\documentclass[12pt]{article}
\usepackage[colorlinks, citecolor=blue,anchorcolor=red,menucolor=red, linkcolor=red,filecolor=red,urlcolor=blue,frenchlinks=red]{hyperref}
\usepackage{multirow}
\usepackage[left=2cm,right=2cm,top=2cm,bottom=2cm]{geometry}
\usepackage{cite}
\usepackage{psfrag}
\usepackage[demo]{graphicx}
\usepackage{graphicx}
\usepackage{float}
\usepackage{graphics}
\usepackage{caption}
\usepackage{epsfig}
\usepackage{color}
\usepackage{booktabs}
\usepackage{amsmath,amsfonts,amssymb}
\usepackage{pstcol,pst-fill,pst-grad}
\usepackage{pstricks,pst-fill,pst-grad}
\usepackage{euscript}
\usepackage{pstricks}
\usepackage{wrapfig}
\usepackage{slashed}
\usepackage[toc,page]{appendix}
\usepackage{here}
\usepackage{array}

\newcommand{\orcid}[1]{\hspace{1mm}\href{https://orcid.org/#1}{\includegraphics[height=0.3cm,keepaspectratio]{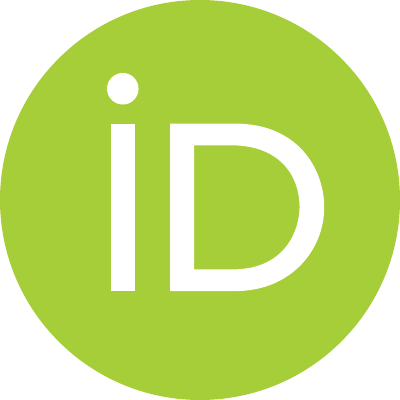}}}
\begin{document}
	\title{\vspace{-3cm}
		\hfill\parbox{4cm}{\normalsize \emph{}}\\
		\vspace{1cm}
		{}}
	\vspace{2cm}
	\author{\thanks{} \\
		{\it {\small}}\\
		{\it {\small }}\\
		{\it {\small}}\\				
	}
	\title{\vspace{-3cm}
		\vspace{1cm}
		{Interplay of elliptical laser polarization and scattering geometry in laser-assisted pair production of doubly charged Higgs bosons in the Higgs Triplet Model}}
	\vspace{2cm}
	
	\author{%
		S. El Handi\textsuperscript{1}%
		\thanks{Corresponding author: s.elhandi@uiz.ac.ma},
		Y. Attaourti\textsuperscript{2},
		B. Taki\textsuperscript{1},
		E. Hrour\textsuperscript{3},
		M. Abaragh\textsuperscript{4}\\[0.8em]
		\begin{minipage}{0.94\textwidth}
			\centering
			\small\itshape
			\textsuperscript{1}\,Laboratory of Theoretical and High Energy Physics (LPTHE),\\
			Faculty of Sciences, Ibnou Zohr University, B.P. 8106, Agadir, Morocco.\\[0.5em]
			\textsuperscript{2}\,Laboratory of High Energy Physics, Astronomy and Space Sciences,\\
			Faculty of Sciences Semlalia, Cadi Ayyad University, B.P. 511, 40000, Morocco.\\[0.5em]
			\textsuperscript{3}\,Superior School of Technology of Beni Mellal,\\
			Research Team in Theoretical Physics and Materials (RTTPM),\\
			Sultan Moulay Slimane University, Beni Mellal 23000, Morocco.\\[0.5em]
			\textsuperscript{4}\,Polydisciplinary Faculty of Beni Mellal,\\
			Research Team in Theoretical Physics and Materials (RTTPM),\\
			Sultan Moulay Slimane University, Beni Mellal 23000, Morocco.
		\end{minipage}%
	}
	\maketitle \setcounter{page}{1}
	\date{}
	\begin{abstract}
		We study the pair production of a doubly charged Higgs boson via $e^{-} e^{+}$ in the presence of a polarized laser field in the Higgs Triplet Model (HTM), mediated by the photon and the $Z$ boson exchange. A unified formulation based on generalized Bessel functions is developed for a monochromatic plane wave with arbitrary ellipticity, encompassing linear, circular, and elliptical polarization states. The numerical analysis shows a strong interplay between laser polarization and scattering geometry, affecting both the characteristic multiphoton domain and the relative strengths of individual harmonic contributions. At a fixed cycle-averaged squared laser potential, the peak ratio normalized to linear polarization exhibits distinct geometry-dependent trends, including a reversal of the polarization hierarchy between different incidence configurations. The dependence on the collision energy and doubly charged Higgs boson mass is also investigated, revealing pronounced kinematic effects on the differential cross section. In contrast, the relative polarization pattern remains similar over a broad mass range. These results provide a unified characterization of polarization and geometrical effects on the multiphoton structure of this process.
	\end{abstract}
	\maketitle
	\section{Introduction}
	\label{sec:intro}
	Over the past few decades, advances in laser technology have led to remarkable progress in the generation and control of electromagnetic fields, with peak laser intensities reaching values of the order of $10^{23} \,\, W/cm^{2}$ \cite{1}. These developments have stimulated considerable theoretical interest in the interaction of charged particles with external electromagnetic fields. In particular, laser-assisted processes have been widely investigated within quantum electrodynamics, providing a framework for studying how the properties of an external field influence particle scattering and decay processes \cite{2,3}. The presence of a laser field introduces multiphoton contributions that depend on the field parameters, including its strength, frequency, and polarization. Understanding the role of these parameters and their interplay with the scattering geometry remains an important aspect of the theoretical description of laser-assisted particle interactions.
	
	Several studies have explored different aspects of particle interactions in external electromagnetic fields. For instance, Ref. \cite{4} investigated the effects of the anomalous magnetic moment in electron-hydrogen elastic collisions in a circularly polarized laser field, confirming the validity of QED predictions in intense electromagnetic fields. In \cite{5}, nonlinear Compton scattering and nonlinear Breit-Wheeler pair production were studied by incorporating particle-state damping effects. Other investigations have extended laser-assisted processes to physics beyond the Standard Model (BSM), including searches for light and invisible particles through nonlinear Compton scattering \cite{6}. In this context, the Volkov state formalism provides an analytic framework for describing charged particles in plane-wave backgrounds  \cite{7}, while further studies have considered more complex laser configurations and strong-field regimes \cite{8,9}.
	
	Since its prediction in the 1960s \cite{10} as a key element in explaining how elementary particles acquire mass through spontaneous electroweak symmetry breaking, the Higgs boson has been the focus of intense theoretical and experimental efforts. For several decades, major experimental programs at CERN and Fermilab \cite{11,12} were dedicated to its search. This longstanding quest culminated on July 4, 2012, when the ATLAS and CMS collaborations at the Large Hadron Collider (LHC) announced the observation of a new scalar particle with a mass of about 125 GeV \cite{13}, consistent with the Higgs boson predicted by the Standard Model (SM). Despite this discovery, many fundamental questions remain unanswered, notably the tiny neutrino masses, the origin of matter–antimatter asymmetry, and the nature of dark matter. Consequently, several extensions of the SM have been proposed to address these issues, such as the Two Higgs Doublet Model (2HDM) \cite{14} and the Minimal Supersymmetric Standard Model (MSSM) \cite{15,16}, which contain an extended Higgs sector and predict additional neutral and charged scalar states ($H$, $A$, $H^\pm$) \cite{17}.  Among the simplest and most motivated extensions addressing the neutrino mass problem \cite{18} is the Higgs Triplet Model (HTM)~\cite{19, 20, 21, 22}, which introduces an $SU(2)_L$ scalar triplet field with hypercharge $Y=2$.  A distinctive feature of this model is the prediction of a doubly charged Higgs boson ($H^{\pm\pm}$), which plays a key role in generating neutrino masses through the Type-II seesaw mechanism~\cite{23}. The search for this particle remains an important objective at the HL-LHC and at proposed future colliders, including the FCC and Muon Collider \cite{24}. The doubly charged Higgs boson can be produced through the Drell-Yan process ($q \bar{q} \rightarrow \gamma^{\star}, Z^{\star} \rightarrow H^{++}H^{--}$) \cite{25}, associated production \cite{26} ($q \bar{q} \rightarrow W^{\pm} \rightarrow H^{\pm\pm}H^{\mp}$), Vector Boson Fusion $W^{\pm}W^{\pm} \rightarrow H^{\pm\pm}$ \cite{27}, and photon-photon fusion $\gamma \gamma \rightarrow H^{++}H^{--}$  \cite{28}. Its main decay channels include $H^{\pm\pm} \rightarrow l^{\pm}l^{\pm}$ or $H^{\pm\pm} \rightarrow W^{\pm} W^{\pm}$  \cite{29,30} .
	
	At the International Linear Collider (ILC), several studies have investigated the production of doubly charged Higgs bosons through $e^{+}e^{-}$ annihilation. In Ref \cite{31}, the authors examined the associated production $e^{+}e^{-} \rightarrow h H^{++}H^{--}$ within the framework of the HTM, where $h$ denotes the SM Higgs boson. Moreover, that work explored the possibility of distinguishing between the HTM and the Minimal Left-Right Symmetric Model (MLRSM) in collider experiments via pair production of $H^{\pm\pm}$ and their subsequent leptonic decays. More recently, the production of doubly charged Higgs bosons at future $e^{+}e^{-}$ colliders has been investigated within the two-Higgs-doublet model with a type-II seesaw mechanism \cite{32}. The laser-assisted production of $H^{\pm\pm}$ through $e^{+}e^{-}$ annihilation was previously investigated in \cite{33} by considering a monochromatic circularly polarized electromagnetic field and a fixed collision geometry. However, the dependence of this process on a continuously varying polarization state, and particularly its interplay with the orientation of the incoming and outgoing particles relative to the laser propagation direction, was not systematically investigated in that analysis.
	
	Motivated by the foregoing discussion, the present work investigates how the polarization state of the laser field and the scattering geometry jointly affect the cross section for the production of doubly charged Higgs bosons through $e^{+}e^{-}$ annihilation. In particular, we employ a general formulation in which the laser polarization is described in a unified way through the ellipticity parameter $\eta$, allowing the linear, circular, and elliptical polarization regimes to be treated within the same framework. This formulation enables us to investigate the continuous dependence of the angular distribution and the photon-number contributions on the laser polarization, as well as their sensitivity to the scattering geometry. The paper is organized as follows. In Section \ref{sec2}, we derive the expression for the differential cross section (DCS) after defining the states of all particles involved in this process and the corresponding couplings for both $ \gamma$- and $Z$-mediated diagrams. We then present the ordinary and generalized Bessel functions that appear in the scattering amplitude, whose arguments depend on the ellipticity parameter $\eta$, through the quantities $z$ and $\alpha_3$. At the same time, $n$ denotes the number of photons exchanged. In Section \ref{sec3}, we discuss the numerical results for the laser-assisted production cross section of the process $e^{+}e^{-}\rightarrow H^{++}H^{--}$, under linear, circular, and elliptical laser polarization for various geometries. Finally, Section \ref{concl} presents our main conclusions. 
	
	We note that throughout this work, we adopt natural units $\hbar = c = 1$, use the metric tensor $g^{\mu\nu}=(1,-1,-1,-1)$, and take the electron charge as $e=-|e|$.  
	
	\section{Theoretical Framework}
	\label{sec2}
	\subsection{HTM in a nutshell}
	Firstly proposed by Magg. et {\it al} \cite{23} to provide a natural explanation for small neutrino masses via the Type-II seesaw mechanism, the HTM extends the SM scalar sector by introducing a complex $SU(2)_L$ triplet scalar field $\Delta$ with hypercharge $Y = 2$.
	
	The most general gauge invariant and renormalizable scalar potential $V(H,\Delta)$ can be expressed as~\cite{34,35},
	\begin{eqnarray}
		V(H,\Delta)&=&-m_H^2{H^\dagger{H}}+\frac{\lambda}{4}(H^\dagger{H})^2+M_\Delta^2Tr(\Delta^{\dagger}{\Delta})\nonumber\\
		&&+\lambda_1(H^\dagger{H})Tr(\Delta^{\dagger}{\Delta})+\lambda_2(Tr\Delta^{\dagger}{\Delta})^2
		+\lambda_3Tr(\Delta^{\dagger}{\Delta})^2\nonumber\\
		&&+\lambda_4{H^\dagger\Delta\Delta^{\dagger}H}+[\mu(H^T{i}\tau_2\Delta^{\dagger}H)+\text{h.c.}]
		\label{higgspot} 
	\end{eqnarray}
	in which both multiplets $H$ and $\Delta$ are given by the following decomposition:  
	\begin{eqnarray}
		H=\left(
		\begin{array}{c}
			\phi^+ \\
			\phi^0 \\
		\end{array}
		\right)\,{\rm and}\,\,  \Delta =\left(
		\begin{array}{cc}
			\delta^+/\sqrt{2} & \delta^{++} \\
			\delta^0 & -\delta^+/\sqrt{2}\\
		\end{array}
		\right)
	\end{eqnarray}
	where $\tau_2$ stands for the Pauli matrix. The parameters $\mu$, $\lambda$, $\lambda_1$--$\lambda_4$, $m^2_{H}$ and $M^2_{\Delta}$ are assumed to be real.
	
	After electroweak symmetry breaking, the HTM scalar spectrum contains the SM-like Higgs boson $h^0$, the doubly charged Higgs states $H^{\pm\pm}$, the singly charged Higgs $H^{\pm}$, and two additional neutral scalars, the CP-even $H^0$ and the CP-odd $A$. From a phenomenological point of view, the presence of the $H^{\pm\pm}$ is one of the most striking signatures of the model, as it directly couples to same-sign lepton pairs or to $W^\pm W^\pm$ gauge bosons, depending on the triplet vacuum expectation value $v_\Delta$. The latter is constrained to be below a few GeV to ensure consistency with precision electroweak data. In the present work, we focus exclusively on the production process $e^+e^- \rightarrow H^{++} H^{--}$  mediated by the $\gamma$ and $ Z$ boson. Within the gauge-mediated production mechanism considered here, no specific values are assumed for the triplet vacuum expectation value $v_{\Delta}$ or the leptonic Yukawa couplings, as these parameters primarily determine the subsequent decay pattern of $H^{\pm\pm}$.  
	\subsection{Laser field: description and formalism}
	The production of doubly charged Higgs boson pairs through electron-positron annihilation provides a useful framework for investigating the extended scalar sector of the HTM. In the present work, we examine how an external electromagnetic field influences the theoretical description of this process, with particular emphasis on the combined effects of laser polarization and scattering geometry. Within the Volkov formalism, the interaction of the incoming electron and positron with the laser field introduces contributions associated with different numbers of photons exchanged. These contributions depend on the particle momenta and field parameters and on the relative orientation of the incident particle momentum and the laser polarization directions. Our analysis focuses on the resulting harmonic structure of the differential cross section in an idealized monochromatic wave background.\\
	The laser field is treated classically and described by a monochromatic, elliptically polarized electromagnetic wave. Its corresponding four-potential is given by :
	\begin{equation}
		A^{\mu} = a_{1}^{\mu}\cos\phi + a_{2}^{\mu}\sin\phi\tan(\eta/2),
		\label{eq1}
	\end{equation}
	with $\phi=k.x$ is the phase of the laser field. The polarization four-vectors $a_{1}$ and $a_{2}$ satisfy the following equations: $a_{1}.a_{2}$=0 and $a_{1}^{2}=a_{2}^{2}=a^{2}=-\lvert\pmb{a}\lvert^2=-(\varepsilon_{0}/\omega)^2$ where $\varepsilon_{0}$ is the electric field strength and $\omega$ is the laser frequency. The Lorentz condition implies that $k.a_{1} = 0$ and $k.a_{2} = 0$. $\eta$ is the degree of ellipticity of the laser field.\\
	In the presence of an electromagnetic field with elliptical polarization, the Dirac-Volkov functions $\psi_{p_{1}, s_{1}}(x)$ and $\psi_{p_{2}, s_{2}}(x)$ describe the electron and positron particles successively :
	\begin{equation}
		\begin{cases}
			\psi_{p_{1}, s_{1}}(x)= \Big[1+\dfrac{e }{2(k.p_{1})}(\slashed k \slashed a_{1}\cos\phi +\slashed k \slashed a_{2}\tan(\eta/2).\sin\phi)\Big] \frac{u(p_{1}, s_{1})}{\sqrt{2Q_{1}V}} \exp^{iS(q_{1},s_{1})}&\\
			\psi_{p_{2}, s_{2}}(x)= \Big[1-\dfrac{e }{2(k.p_{2})}(\slashed k \slashed a_{1}\cos\phi +\slashed k \slashed a_{2}\tan(\eta/2).\sin\phi)\Big] \frac{v(p_{2}, s_{2})}{\sqrt{2Q_{2}V}} \exp^{iS(q_{2},s_{2})},
		\end{cases}
		\label{eq2}
	\end{equation}
	where
	\begin{equation}
		\begin{cases}
			S(q_{1},s_{1})=- q_{1}x -\frac{e(a_{1}.p_{1})}{k.p_{1}}\sin\phi +\frac{e^{2}a^{2}}{8(k.p_{1})}(1-\tan^{2}(\eta/2))\sin2\phi+\frac{e(a_{2}.p_{1})}{k.p_{1}}\tan(\eta/2)\cos\phi&\\
			S(q_{2},s_{2})=+ q_{2}x -\frac{e(a_{1}.p_{2})}{k.p_{2}}\sin\phi -\frac{e^{2}a^{2}}{8(k.p_{2})}(1-\tan^{2}(\eta/2))\sin2\phi+ \frac{e(a_{2}.p_{2})}{k.p_{2}}\tan(\eta/2)\cos\phi
		\end{cases}
		\label{eq3}
	\end{equation}
	Here $x$ is the space-time coordinate of the incoming particles. The Dirac bispinor of the electron and positron are represented by $u(p_{1}, s_{1})$ and $v(p_{2}, s_{2})$, respectively, where $s_i\, (i = 1, 2)$ denote their spins and satisfy the sum-rules: $\sum_{s}u(p_{1}, s_{1})\bar{u}(p_{1},s_{1})=(\slashed p_{1}+m_{e})$ and $\sum_{s}v(p_{2}, s_{2})\bar{v}(p_{2}, s_{2})=(\slashed p_{2}-m_{e})$. The Feynman slash notation is used \cite{36} ($ \slashed a_{1}=\gamma^{\mu}a_{1}$).\\
	The quasi-momentum of the electron and positron dressed by the laser field becomes: 
	\begin{equation}
		q^{\mu}_{i}\,(i=1, 2) =(Q_ {i},\bold{q}_{i})=p^{\mu}_{i} -\dfrac{e^{2}a^{2}}{4(k.p_{i})} (1+\tan^{2}(\eta/2))k^{\mu}
		\label{eq4}
	\end{equation}
	The effective mass acquired by the dressed incoming particles is such that:
	\begin{equation}
		q_{i}^{2}=m_{e}^{*2} = m_{e}^{2} - \dfrac{e^{2}a^{2}}{2}(1+\tan^{2}(\eta/2)),
		\label{eq5}
	\end{equation}
	where $m_e$ is the mass of the incoming particles. 
	
\begin{figure}[H]
		\centering
		\includegraphics[scale=0.2]{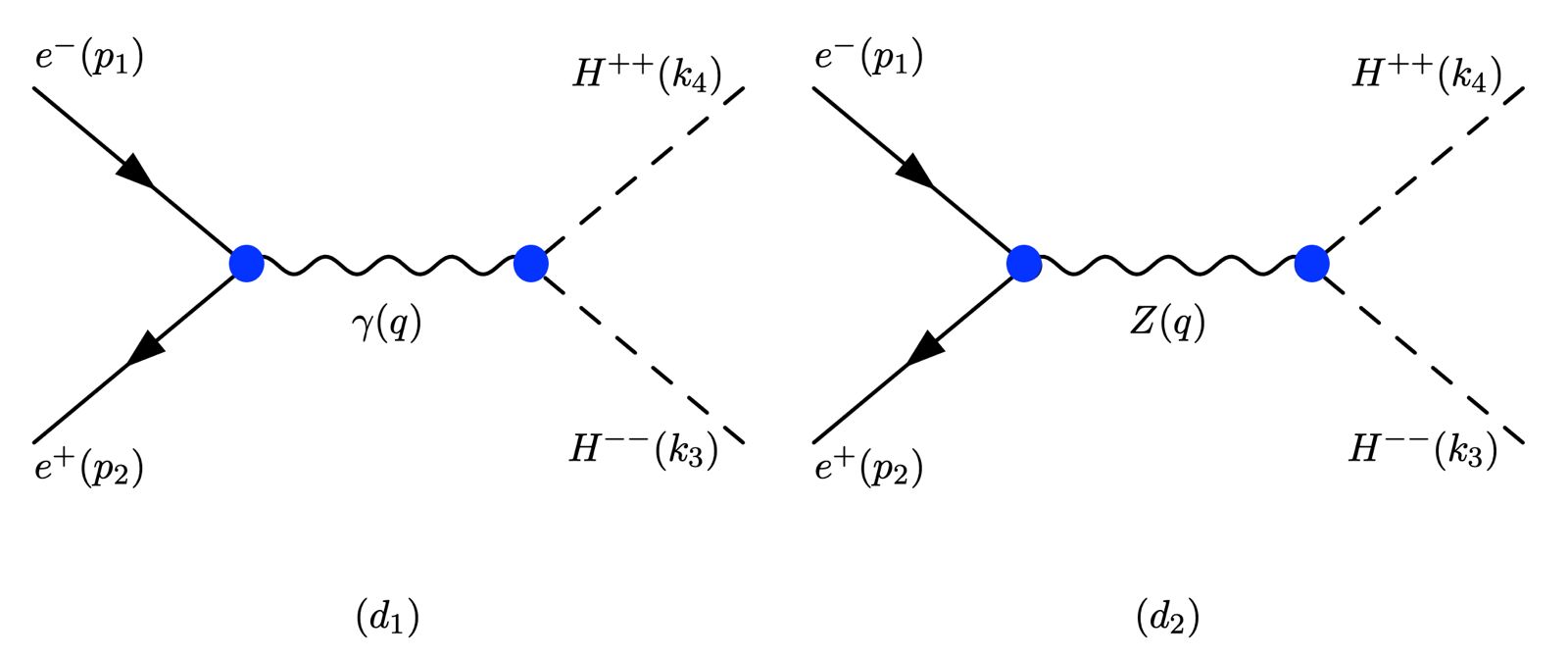}\hspace*{0.4cm}     
		\caption{Feynman diagrams of the process $e^{+}e^{-}\rightarrow H^{++}H^{--}$.}
		\label{fig1}       
	\end{figure}

	Following the approximation adopted in Ref.~\cite{33}, the outgoing doubly charged Higgs bosons $H^{++}$ and $H^{--}$ are described by free Klein-Gordon states. This choice is motivated by the large masses of the produced Higgs bosons, for which laser-induced effective-mass corrections are negligible under the field conditions considered. Accordingly, our analysis focuses on initial-state laser dressing, neglecting the direct laser-dependent contributions associated with the outgoing charged scalar states. Their wave functions are given by :
	\begin{equation}
		\begin{aligned}
			\varphi_{3}(y) &=\dfrac{1}{\sqrt{2 E_{3} V}} e^{-ik_{3}y},\\[2mm]
			\varphi_{4}(y) &=\dfrac{1}{\sqrt{2 E_{4} V}} e^{-ik_{4}y}
			\label{eq6}
		\end{aligned}
	\end{equation}
	where $E_{3}$ and $E_{4}$ are respectively the energies of the $H^{--}$  and $H^{++}$ Higgs bosons and $k_{3}$ and $k_{4}$ indicate their free four-momentum and $y$ is their corresponding space time coordinate.\\
	Having specified the particle states, we now consider the process $e^{+}e^{-}\rightarrow H^{++}H^{--}$ mediated by $\gamma$ and $Z$ boson exchange. The corresponding Feynman diagrams are shown in Fig.~\ref{fig1},
	where the blue blobs refer here to either quantum electrodynamics ($\gamma-e-e$), electroweak ($Z-e-e$), and $H^{\pm\pm}$ gauge couplings within the HTM \cite{35}, these vertices can be written as :
	\begin{equation}
		\gamma H^{++}H^{--} = 2ie(k_{4}-k_{3})^{\nu},\quad\qquad Z H^{++}H^{--} = i\frac{e(1-2\sin({\theta_W})^{2})}{\sin({\theta_W})\cos({\theta_W})}(k_{4}-k_{3})^{\nu}.
		\label{eq7}
	\end{equation}
	
	\begin{equation}
		\gamma e^{-}e^{+} = -ie \gamma^{\mu},\quad\qquad Z e^{-}e^{+} =  \frac{ie}{2\cos({\theta_W}) \sin({\theta_W})} \gamma^{\mu} (g_v^{e} -g_a^{e}\gamma^{5}),
		\label{eq8}
	\end{equation}
	
	with $\theta_W$ is the Weinberg angle $\theta_W=28.75^{\circ}$. For more simplification we denote $\cos({\theta_W})$ and $\sin({\theta_W})$ respectively by $C_W$ and $S_W$. The vector and axial coupling constants are respectively  $g_v^{e}=\dfrac{-1}{2} + 2sin^{2}({\theta_W})$ and $g_a^{e}=-\dfrac{1}{2}$ \cite{36}.
	The Feynman propagators for $\gamma$ and $Z$-boson 
	using the Feynman gauge \cite{38,39} , are given by:
	
	\begin{equation}
		D_{\mu\nu}(x-y)=\int \dfrac{d^{4}q}{(2\pi)^4} \frac{-ig_{\mu\nu} e^{-iq(x-y)}}{q^{2}+i\epsilon},
		\label{eq9}
	\end{equation}
	
	\begin{equation}
		G_{\mu\nu}(x-y)=\int \dfrac{d^{4}q}{(2\pi)^4} \frac{-ig_{\mu\nu} e^{-iq(x-y)}}{q^{2}-M_{Z}^{2}} 
		\label{eq10}
	\end{equation}
	
	where $q$ is the four-momentum and $M_{Z}$ is the rest mass of Z-boson.
	
	The scattering-matrix element of the pair production of doubly charged Higgs bosons via electron-positron annihilation in the presence of the laser field in HTM can be written as: 
	\begin{equation}
		S_{fi}({e}^{+}{e}^{-} \rightarrow H^{++}H^{--})= S_{fi}(\gamma)+S_{fi}(Z)
		\label{eq11}
	\end{equation}

	Subsequently, we substitute the expressions of wave functions and Feynman propagators and the gauge couplings into each matrix element; we obtain:
	\begin{eqnarray}
		S_{fi}(\gamma)&=&\dfrac{-2ie^{2}}{(2\pi)^4}\dfrac{1}{4V^{2}}\dfrac{1}{\sqrt{Q_{1}Q_{2}E_{3}E_{4}}}\int d^{4}x\int d^{4}y\int \dfrac{d^{4}q }{q^{2}+i\epsilon} \nonumber \\ & \times & e^{i(\alpha_{1}\sin\phi+\alpha_{2}\cos\phi+\alpha_{3}\sin2\phi)}  e^{-i(q_{1}+q_{2}+q)x} e^{i(k_{4}+k_{3}+q)y} \nonumber \\ & \times & \bar{v}(p_{2},s_{2})  \Big[1-\dfrac{e }{2(k.p_{2})}(\slashed a_{1}\slashed k cos\phi +\slashed a_{2}\slashed k \tan(\eta/2)\sin\phi)\Big] \gamma^ {\mu}  \nonumber \\ &  & \Big[1+\dfrac{e }{2(k.p_{1})}(\slashed k \slashed a_{1}\cos\phi +\slashed k \slashed a_{2}\tan(\eta/2)\sin\phi)\Big] u(p_{1},s_{1})(k_{4}-k_{3})_{\mu}
		\label{eq12}  
	\end{eqnarray}
	and
	\begin{eqnarray}
		S_{fi}(Z)&=&\dfrac{1}{(2\pi)^4}\dfrac{1}{4V^{2}}\dfrac{1}{\sqrt{Q_{1}Q_{2}E_{3}E_{4}}}\int d^{4}x\int d^{4}y\int \dfrac{d^{4}q }{q^{2}-M^{2}_{Z}} \nonumber \\ & \times & e^{i(\alpha_{1}\sin\phi+\alpha_{2}\cos\phi+\alpha_{3}\sin2\phi)} e^{-i(q_{1}+q_{2}+q)x} e^{i(k_{4}+k_{3}+q)y} \nonumber \\ & \times & \frac{ie^{2}}{2C_{W}S_{W}} \frac{(1-2S_{W}^{2})}{S_{W}C_{W}} \nonumber \\ & \times & \bar{v}(p_{2},s_{2})  \Big[1-\dfrac{e }{2(k.p_{2})}(\slashed a_{1}\slashed k cos\phi +\slashed a_{2}\slashed k \tan(\eta/2)\sin\phi)\Big] \gamma^ {\mu}(g_v^{e}-g_a^{e} \gamma^ {5}) \nonumber \\ &  & \Big[1+\dfrac{e }{2(k.p_{1})}(\slashed k \slashed a_{1}\cos\phi +\slashed k \slashed a_{2}\tan(\eta/2)\sin\phi)\Big] u(p_{1},s_{1})(k_{4}-k_{3})_{\mu},
		\label{eq13}  
	\end{eqnarray}
	whereas the quantities $\alpha_{1}$, $\alpha_{2}$ and $\alpha_{3}$ are given by:
	\begin{equation}
		\alpha_{1}=e\Bigg(\dfrac{(a_{1}.p_{2})}{(k.p_{2})}-\dfrac{(a_{1}.p_{1})}{(k.p_{1})}\Bigg),  \qquad \alpha_{2}=e\Bigg(\dfrac{(a_{2}.p_{1})}{(k.p_{1})}-\dfrac{(a_{2}.p_{2})}{(k.p_{2})}\Bigg)tan(\eta/2),
		\label{eq14} 
	\end{equation}
	\begin{equation*}
		\alpha_{3}=\dfrac{e^{2}a^{2}}{8}(1-tan^{2}(\eta/2))\Bigg(\dfrac{1}{k.p_{1}}+\dfrac{1}{k.p_{2}}\Bigg)
	\end{equation*}
	We now introduce the well-known ordinary and generalized Bessel functions \cite{40,41} defined by:
	\begin{equation}
		e^{i(\alpha_{1}\sin\phi+\alpha_{2}\cos\phi+\alpha_{3}\sin2\phi)}= \sum_{n=-\infty}^{n=+\infty} \Bigg( \sum_{l=-\infty}^{l=+\infty} J_{l}(\alpha_{3})J_{n-2l}(z)e^{i(n-2l)\phi_{0}} \Bigg)e^{in\phi}
		\label{eq15}
	\end{equation}
	with $z$ denotes the argument of the Bessel function and $\phi_0$ its phase, taken as:
	\begin{equation}
		z=\sqrt{\alpha_{1}^{2}+\alpha_{2}^{2}} \quad,\quad\qquad \phi_{0}=\arctan \left(\frac{\alpha_{2}}{\alpha_{1}}\right),
		\label{eq16}
	\end{equation}
	and $n$ is the number of photons exchanged between the incoming particles and the laser field. \\
	This decomposition naturally incorporates the ellipticity of the laser field and allows a unified treatment of all polarization states.\\
	
	We apply the following transformation :
	\begin{equation}
		\begin{split}
			\begin{Bmatrix}
				1\\
				\cos(\phi)\\
				\sin(\phi) \\
				\cos^{2}(\phi)
			\end{Bmatrix} 
			e^{i(\alpha_{1}\sin\phi+\alpha_{2}\cos\phi+\alpha_{3}\sin2\phi)}  =\sum_{n=-\infty}^{+\infty} \begin{Bmatrix}
				B_{n}(\alpha_{3},z)\\
				B_{1n}(\alpha_{3},z)\\
				B_{2n}(\alpha_{3},z)\\
				B_{3n}(\alpha_{3},z)
			\end{Bmatrix}
			~e^{i n \phi},
		\end{split}
		\label{eq17} 
	\end{equation} 
	where $B_{n}(\alpha_{3},z)$, $B_{1n}(\alpha_{3},z)$,$B_{2n}(\alpha_{3},z)$ and $B_{3n}(\alpha_{3},z)$ are given in terms of ordinary Bessel function \cite{40}:
	\begin{equation}
		\begin{split}
			B_{n}(\alpha_{3},z) &=  \sum_{l=-\infty}^{l=+\infty} J_{l}(\alpha_{3})J_{n-2l}(z)e^{i(n-2l)\phi_{0}} ,\\
			B_{1n}(\alpha_{3},z) &=\sum_{l=-\infty}^{l=+\infty} \dfrac{1}{2}J_{l}(\alpha_{3})\Big(J_{n-1-2l}(z)e^{i(n-1-2l)\phi_{0}}+J_{n+1-2l}(z)e^{i(n+1-2l)\phi_{0}}\Big)  ,\\
			B_{2n}(\alpha_{3},z) &=\sum_{l=-\infty}^{l=+\infty} \dfrac{1}{2i}J_{l}(\alpha_{3})\Big(J_{n-1-2l}(z)e^{i(n-1-2l)\phi_{0}}-J_{n+1-2l}(z)e^{i(n+1-2l)\phi_{0}}\Big) ,\\
			B_{3n}(\alpha_{3},z)&=\sum_{l=-\infty}^{l=+\infty} J_{l}(\alpha_{3})\Bigg(\dfrac{1}{2}J_{n-2l}(z)e^{i(n-2l)\phi_{0}}+\dfrac{1}{4}J_{n+2-2l}(z)e^{i(n+2-2l)\phi_{0}}+\dfrac{1}{4}J_{n-2-2l}(z)e^{i(n-2-2l)\phi_{0}}\Big) 
		\end{split}
		\label{eq18} 
	\end{equation} 
	Using the generalized Bessel function allows us to derive a unified expression for the DCS, valid for arbitrary laser polarization. In this framework, the polarization state is entirely characterized by the ellipticity parameter $\eta$, which enables a continuous transition between linear, circular, and elliptical polarizations within a single formalism.\\
	This approach avoids the need for separate calculations for each polarization case and provides a consistent and systematic description of laser-assisted scattering processes.\\
	Thus, we get the expression of the S-matrix element as:
	\begin{eqnarray}
		S_{fi}(\gamma)&=&\sum_{n=-\infty}^{+\infty}\dfrac{1}{4V^{2}}\dfrac{(2\pi)^4~\delta^{4}(k_{4}+k_{3}-q_{1}-q_{2}+ n k)}{\sqrt{Q_{1}Q_{2}E_{3}E_{4}}} \dfrac{-2ie^{2} }{(q_{1}+q_{2} -nk)^{2}} \nonumber \\ & \times & \bar{v}(p_{2},s_{2})  \Gamma_{n}(\gamma) u(p_{1},s_{1})(k_{4}-k_{3})_{\mu}
		\label{eq19}  
	\end{eqnarray}
	and
	\begin{eqnarray}
		S_{fi}(Z)&=&\sum_{n=-\infty}^{+\infty}\dfrac{1}{4V^{2}}\dfrac{(2\pi)^4~\delta^{4}(k_{4}+k_{3}-q_{1}-q_{2}+ n k)}{\sqrt{Q_{1}Q_{2}E_{3}E_{4}}}\frac{ie^{2}}{2C_{W}S_{W}}    \frac{(1-2S_{W}^{2})}{S_{W}C_{W}} \nonumber \\ & \times & \dfrac{1}{(q_{1}+q_{2} -nk)^{2}-M^{2}_{Z}} \bar{v}(p_{2},s_{2})  \Gamma_{n}(Z) u(p_{1},s_{1})(k_{4}-k_{3})_{\mu}
		\label{eq20}  
	\end{eqnarray}
	with $\Gamma_{n}(Z)$ and  $\Gamma_{n}(\gamma)$ are given by:
	\begin{eqnarray}
		\Gamma_{n}(\gamma)&=&\xi_{0}B_{n}(\alpha_{3},z)+\xi_{1}B_{1n}(\alpha_{3},z)+\xi_{2}B_{2n}(\alpha_{3},z)+\xi_{3}B_{3n}(\alpha_{3},z)
		\label{eq21}
	\end{eqnarray}
	\begin{eqnarray}
		\Gamma_{n}(Z)&=&\chi_{0}B_{n}(\alpha_{3},z)+\chi_{1}B_{1n}(\alpha_{3},z)+\chi_{2}B_{2n}(\alpha_{3},z)+\chi_{3}B_{3n}(\alpha_{3},z)
		\label{eq22}
	\end{eqnarray}
	The coefficients in equations \ref{eq21} and \ref{eq22}, using the notation $C_{i=1,2}=e/2(k.p_{i})$, are expressed as: 
	\begin{equation}
		\begin{cases}
			\xi_{0}=\gamma^{\mu}+2C_{1}C_{2}a^{2}k^{\mu}\slashed k \tan^{2}(\eta/2) &\\ 
			\xi_{1}=C_{1}\gamma^{\mu}\slashed k\slashed a_{1}-C_{2}\slashed a_{1}\slashed k \gamma^{\mu} &\\   
			\xi_{2}=\big(C_{1}\gamma^{\mu}\slashed k\slashed a_{2}-C_{2}\slashed a_{2}\slashed k \gamma^{\mu}\big)\tan(\eta/2) &\\ 
			\xi_{3}=2C_{1}C_{2}a^{2}k^{\mu}\slashed k \big(1-\tan^{2}(\eta/2)\big) &\\ 
			\chi_{0}=\gamma^{\mu}(g_{v}^{e}-g_{a}^{e}\gamma^{5})+2C_{1}C_{2}a^{2}k^{\mu}\slashed k(g_{v}^{e}-g_{a}^{e}\gamma^{5})\tan^{2}(\eta/2)   &\\ 
			\chi_{1}=C_{1}\gamma^{\mu}(g_{v}^{e}-g_{a}^{e}\gamma^{5})\slashed k\slashed a_{1}-C_{2}\slashed a_{1}\slashed k \gamma^{\mu}(g_{v}^{e}-g_{a}^{e}\gamma^{5})   &\\ 
			\chi_{2}=\big(C_{1}\gamma^{\mu}(g_{v}^{e}-g_{a}^{e}\gamma^{5})\slashed k\slashed a_{2}-C_{2}\slashed a_{2}\slashed k \gamma^{\mu}(g_{v}^{e}-g_{a}^{e}\gamma^{5})\big)\tan(\eta/2) &\\
			\chi_{3}=2C_{1}C_{2}a^{2}k^{\mu}\slashed k (g_{v}^{e}-g_{a}^{e}\gamma^{5})\big(1-\tan^{2}(\eta/2)\big) 
		\end{cases}
		\label{eq23}
	\end{equation}
	As previously mentioned, the S-matrix element of our process is given by the sum of the contributions arising from $\gamma$ and $Z$ coupling, such as: 
	\begin{eqnarray}
		S_{fi}({e}^{+}{e}^{-} \rightarrow H^{++}H^{--})&=&\sum_{n=-\infty}^{+\infty}\dfrac{1}{4V^{2}}\dfrac{(2\pi)^4~\delta^{4}(k_{4}+k_{3}-q_{1}-q_{2}+ n k)}{\sqrt{Q_{1}Q_{2}E_{3}E_{4}}} \big|M_{fi}^{n}\big|,
		\label{eq24}  
	\end{eqnarray}
	where $|M_{fi}|$ is the total scattering amplitude, whose explicit expression is obtained by evaluating the traces using the symbolic computation package FeynCalc \cite{42}. The momentum transfer is $q=q_{1}+q_{2}- n k$.
	
	Within the conventional framework of QED calculations \cite{36,43}, the differential cross section DCS in the center-of-mass frame is obtained using the following expression :
	\begin{equation}
		d\sigma=\dfrac{|S_{fi}({e}^{+}{e}^{-} \rightarrow H^{++}H^{--})|^{2}}{VT}\frac{1}{|J_{inc}|}\frac{1}{\rho}V\int_{}\dfrac{d^{3}k_{3}}{(2\pi)^3}V\int_{}\dfrac{d^{3}k_{4}}{(2\pi)^3},
		\label{25}
	\end{equation}
	with $ |J_{inc}|=\sqrt{(q_{1}q_{2})^2-m_{e}^{*4}}/Q_{1}Q_{2}V$ is the current of the incoming particles and $\rho=1/V$ is the density of particles.
	By averaging over the initial spins and summing over the final ones, and after some algebraic calculations, the differential cross section will be as follows:
	\begin{eqnarray} 
		d\sigma&=&\sum_{n=-\infty}^{+\infty} \dfrac{1}{16\sqrt{(q_{1}q_{2})^2-m_{e}^{*^{4}}}}   \big|\overline{M_{fi}} \big|^{2} \int_{}\dfrac{|\mathbf{k}_{3}|^{2}d|\mathbf{k}_{3}|d\Omega}{(2\pi)^2E_{3}} \int_{}\dfrac{d^{3}k_{4}}{ E_{4}} \nonumber \\  &\times & \delta^{4}(k_{4}+k_{3}-q_{1}-q_{2}+nk).
		\label{26}
	\end{eqnarray}
	The differential cross section corresponding to the doubly charged Higgs boson pair production via $e^{+}e^{-}$ annihilation is given by:
	\begin{eqnarray}
		\dfrac{d\sigma_{n}}{d\Omega}({e}^{+}{e}^{-}\rightarrow H^{++}H^{--})&=&\dfrac{1}{16(2\pi)^2\sqrt{(q_{1}q_{2})^2-m_{e}^{*^{4}}}}  \big|\overline{M_{fi}} \big|^{2} \dfrac{f(x)}{|g^{'}(|\mathbf{k}_{3}|)|_{g(|\mathbf{k}_{3}|)=0}}
		\label{27}
	\end{eqnarray}
	Using the following relation \cite{40}:
	\begin{equation}
		\int d\mathbf x f(\mathbf x) \delta(g(\mathbf x))=\dfrac{f(\mathbf x)}{|g^{'}(\mathbf x)|_{g(\mathbf x)=0}}
		\label{eq28}
	\end{equation}
	where
	\begin{eqnarray}
		\big|g^{'}(|\textbf{k}_{3}|)\big|&=&\dfrac{|\textbf k_{3}|}{\sqrt{|\textbf k_{3}|^{2}+M_{H^{\pm\pm}}^{2}}}\Bigg[\dfrac{e^{2}a^{2}}{2}\big(1+\tan^{2}(\eta/2)\big)\big(\dfrac{1}{k.p_{1}}+\dfrac{1}{k.p_{2}}\big)\omega-2E_{1}-2E_{2}+2n\omega \Bigg]\nonumber \\ &  & -\Bigg[2n+\dfrac{e^{2}a^{2}}{2}\big(1+\tan^{2}(\eta/2)\big)\big(\dfrac{1}{k.p_{1}}+\dfrac{1}{k.p_{2}}\big)\Bigg]\omega\cos\theta_{f}
		\label{29}
	\end{eqnarray}
	with $M_{H^{\pm\pm}}$ is the mass of the doubly charged Higgs Boson.
	
	\section{Results and discussion}
	\label{sec3}
	
\begin{figure}[H]
		\centering
		\includegraphics[scale=0.4]{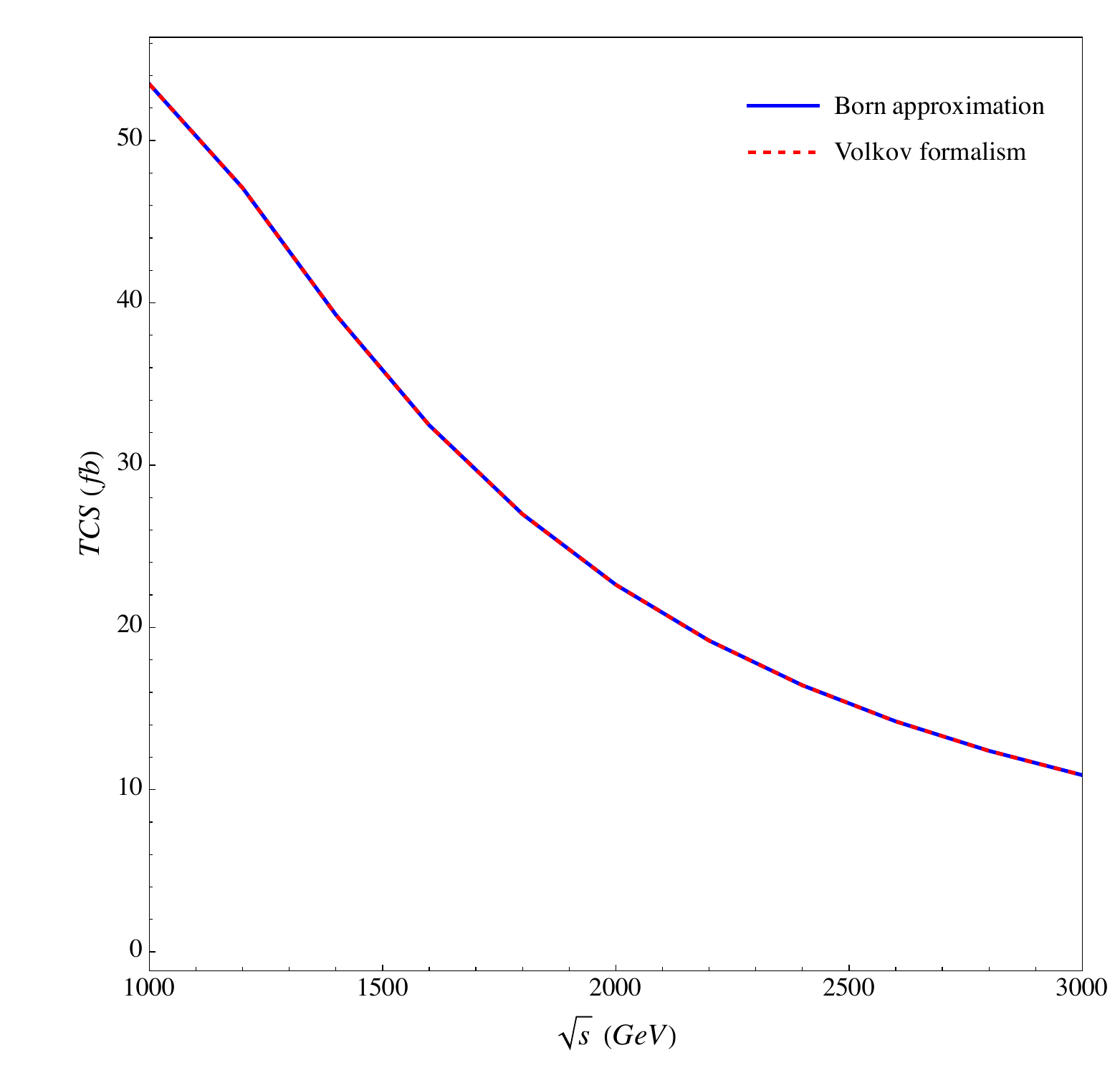}
		\caption{The total cross section (TCS) of the production of $H^{++}H^{--}$  via $e^{+}e^{-}$ annihilation as a function of $\sqrt{s}$. The solid blue curve represents the analytic Born approximation in the absence of a laser field, while the red dashed curve corresponds to the generalized Volkov formalism in the laser-free limit $\varepsilon_{0} = 0\, V.cm^{-1}$ with $ n = 0$. The doubly charged Higgs boson mass is fixed at $M_{H^{\pm\pm}}=300 \,\, GeV$.}
		\label{fig2}
	\end{figure}

	Before presenting the numerical results, we first specify the scattering geometry, in which the direction of the laser propagation is along $Oz$ axis $k=(\omega,0,0,\omega)$, and the polarization four-vectors $a_{1}$ and $a_{2}$ are respectively along $Ox$ and $Oy$ axis such that $a_{1}=(0,\lvert a \rvert,0,0)$ and $a_{2}=(0,0,|a|,0)$, therefore we deduce the average squared potential $\bar{A^{2}}=a^{2}\big(1+\tan^{2}(\eta/2) \big)/2$. Consequently, for $\eta=0$ we have $\bar{A^{2}}=a^{2}/2$ corresponding to linear polarization and for $\eta=\pi/2$ we have $\bar{A^{2}}=a^{2}$ corresponding to circular polarization. We extend the description of the geometry to the framework of spherical coordinates, where the four-momentum of the incoming electron and positron are defined, respectively, as :
	\begin{align*}
		p_{1}&=\bigl(E_{1},|\mathbf p_1|\sin\theta_{i}\cos\phi_{i}, |\mathbf p_1|\sin\theta_{i}\sin\phi_{i}, |\mathbf p_1|\cos\theta_{i}\bigr),\\[4pt]
		p_{2}&=\bigl(E_{2},-|\mathbf p_2|\sin\theta_{i}\cos\phi_{i}, -|\mathbf p_2|\sin\theta_{i}\sin\phi_{i}, -|\mathbf p_2|\cos\theta_{i}\bigr).
	\end{align*}
	For the outgoing doubly charged Higgs boson, we have 
	\begin{align*}
		k_{3}&=\bigl(E_{3},|\mathbf k_3|\sin\theta_{f}\cos\phi_{f},|\mathbf k_3|\sin\theta_{f}\sin\phi_{f}, |\mathbf k_3|\cos\theta_{f}\bigr)
	\end{align*}
	This parametrization enables us to systematically investigate how the laser polarization and the relative orientation of the particle momenta affect the laser-assisted production process.\\
	To validate our theoretical formalism and numerical calculations, we first examine the laser-free limit by comparing the total cross section TCS obtained from the analytical Born approximation with that calculated using our generalized Volkov formalism. For this comparison, we consider a reference incident collision geometry defined by $\theta_{i}=90^{\circ}$, $\phi_{i}=0^{\circ}$, and consider the laser field-free limit with no photons exchanged. As shown in Fig.~\ref{fig2}, the two predictions are practically indistinguishable over the entire centre-of-mass energy range considered. In particular, at $\sqrt{s}=1000 \,\, GeV$, the Born approximation yields a TCS of $53.4864\,\, fb$, while the Volkov formalism gives $53.4862\,\, fb$. This excellent agreement demonstrates that our generalized formalism correctly reproduces the conventional Born result in the absence of an external laser field, providing an independent consistency check before investigating the combined effects of laser polarization and scattering geometry.\\
	
\begin{figure}[H]
		\centering
		\includegraphics[scale=0.45]{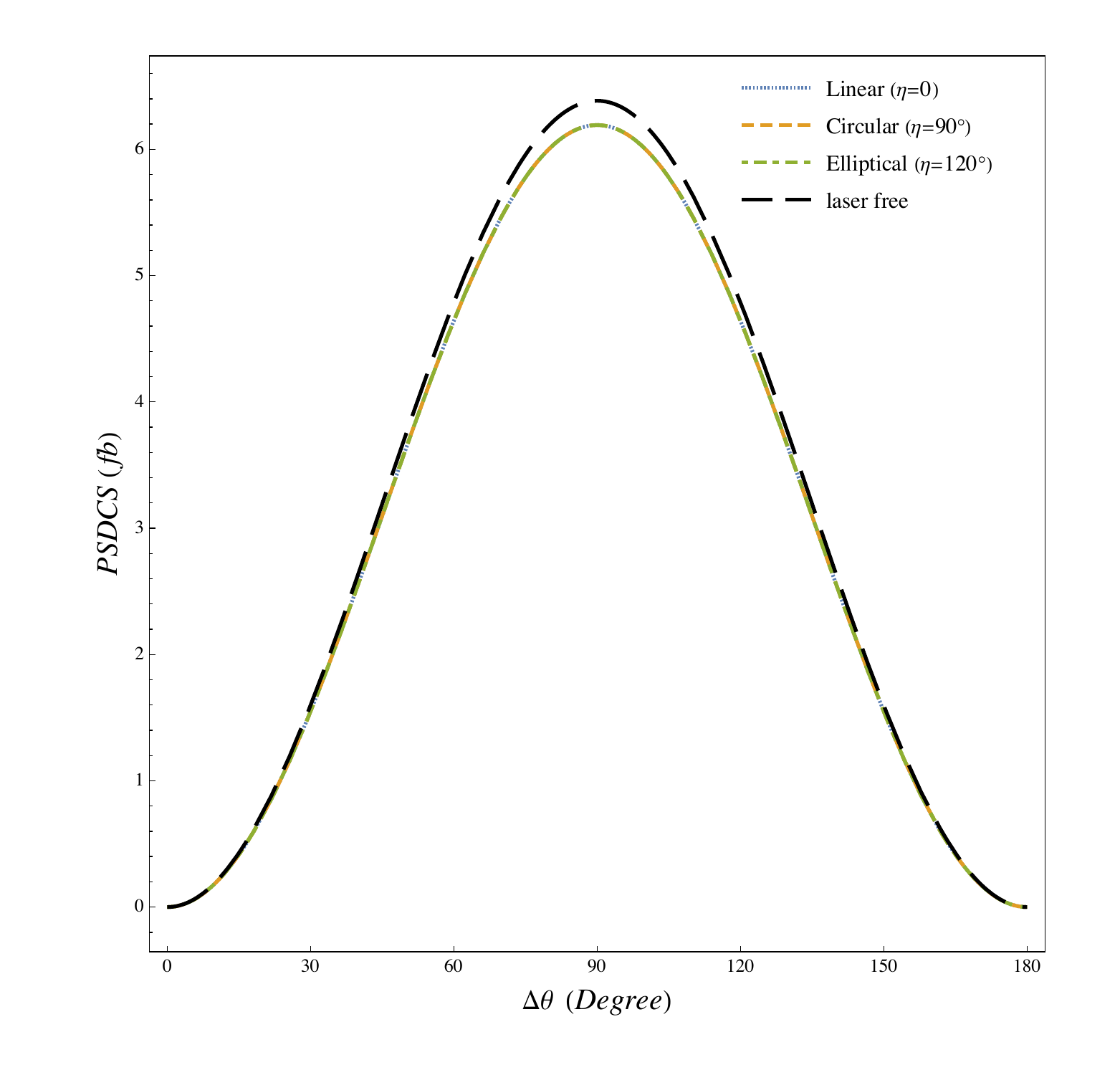}
		\caption{The PSDCS as a function of scattering angle $\Delta \theta=\phi_{f}-\phi_{i}$ for different laser polarization states, with $\theta_{i}= 90^{\circ}$, $\phi_{i}= 0^{\circ}$ and $\theta_{f}=90^{\circ}$, and $-3 \leq n \leq 3$. The laser field strength and frequency are fixed at $\varepsilon=10^{5} \,\,\mathrm{V.cm^{-1}}$ and $\omega=1.17 \,\,\mathrm{eV}$, respectively. The doubly charged Higgs boson mass and the centre-of-mass energy are fixed at $M_{H^{\pm\pm}}=300\,\mathrm{GeV}$ and $\sqrt{s}=1000\,\mathrm{GeV}$, respectively.}
		\label{fig3}
	\end{figure}

\begin{figure}[H]
		\centering
		\includegraphics[scale=0.35]{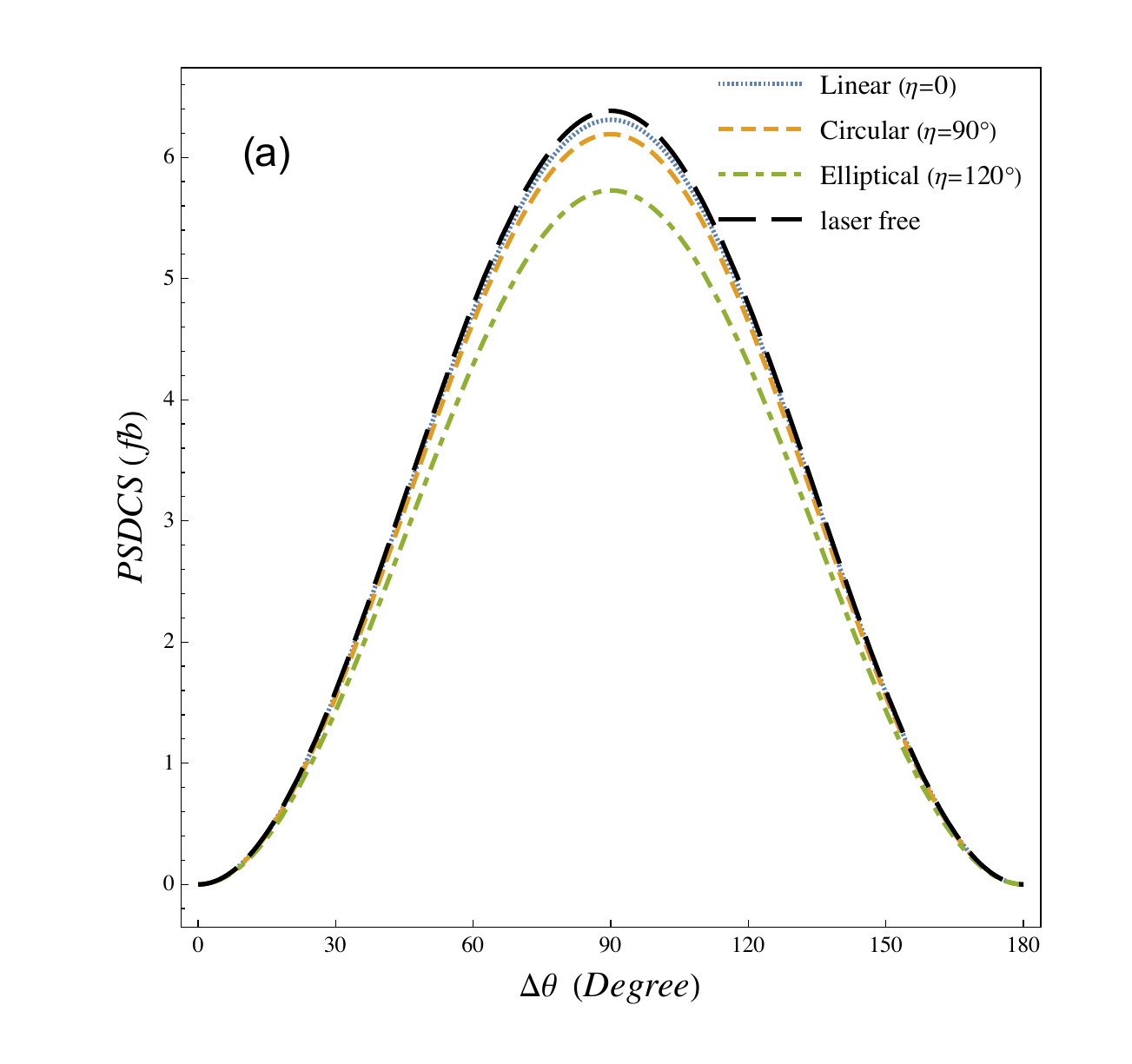}\hspace*{0.1cm}
		\includegraphics[scale=0.35]{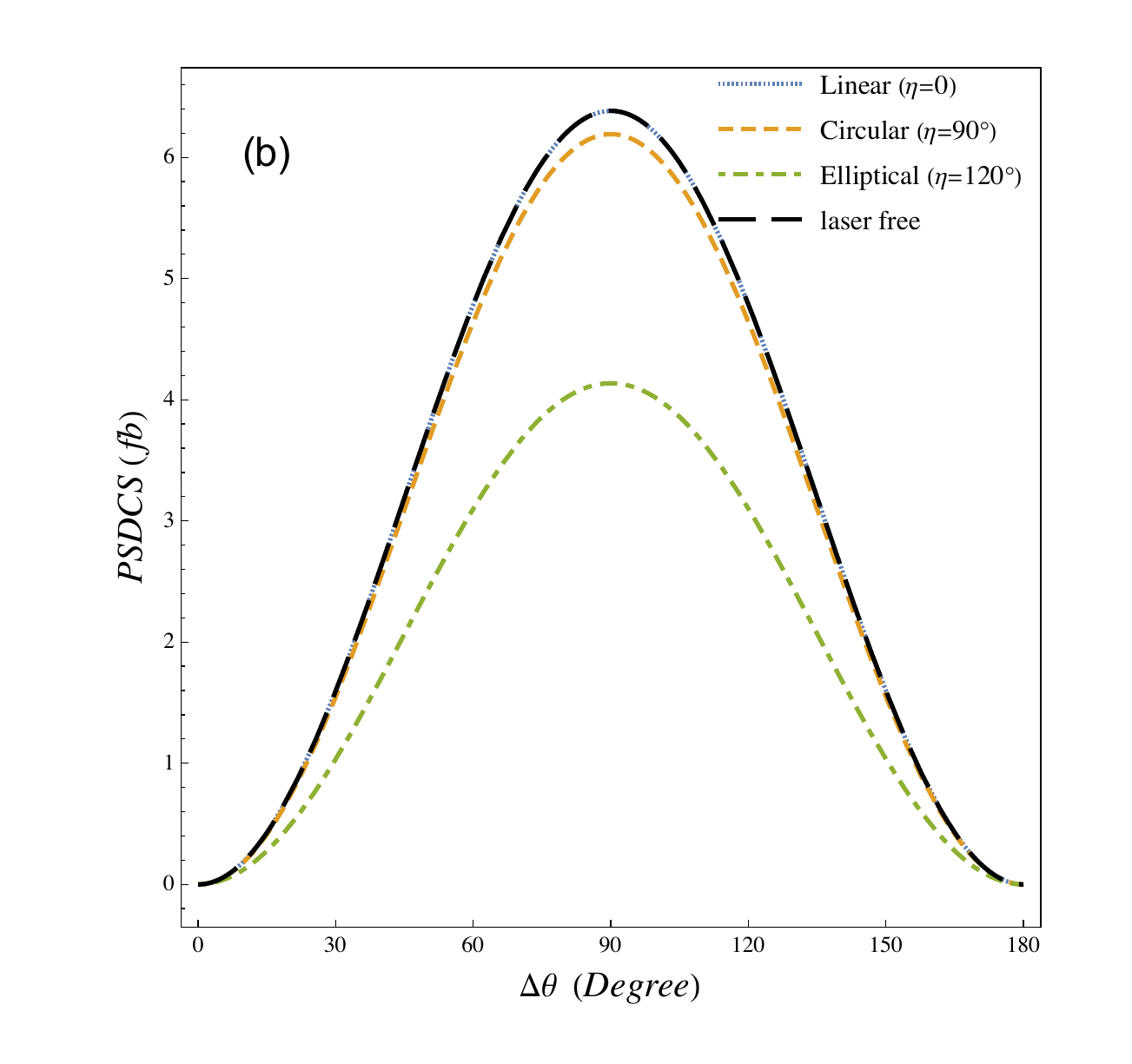}\hspace*{0.1cm}
		\caption{ The PSDCS as a function of scattering angle $\Delta \theta=\phi_{f}-\phi_{i}$. The laser field strength and frequency are fixed at $\varepsilon=10^{5} \,\, \mathrm{V.cm^{-1}}$ and $\omega=1.17 \,\, \mathrm{eV}$, respectively, with $-3 \leq n \leq 3$. The doubly charged Higgs boson mass and the centre-of-mass energy are fixed at $M_{H^{\pm\pm}}=300\,\mathrm{GeV}$ and $\sqrt{s}=1000\,\mathrm{GeV}$, respectively. $\theta_{i}= 90^{\circ}$, $\theta_{f}=90^{\circ}$ in (a): $\phi_{i}=30^{\circ}$ in (b): $\phi_{i}=60^{\circ}$.}
		\label{fig4}
	\end{figure}

\begin{figure}[H]
		\centering
		\includegraphics[scale=0.3]{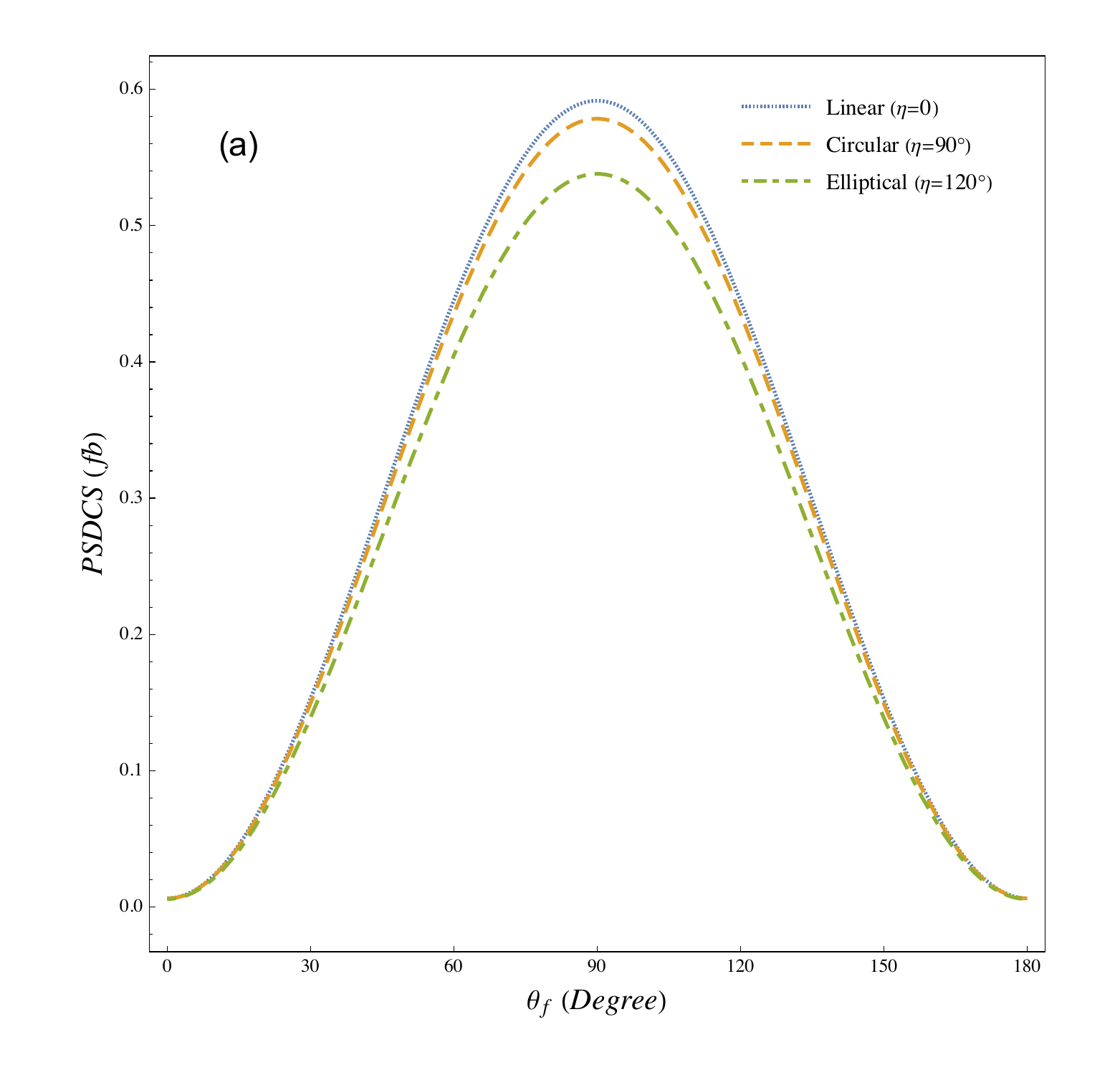}\hspace*{0.1cm}
		\includegraphics[scale=0.3]{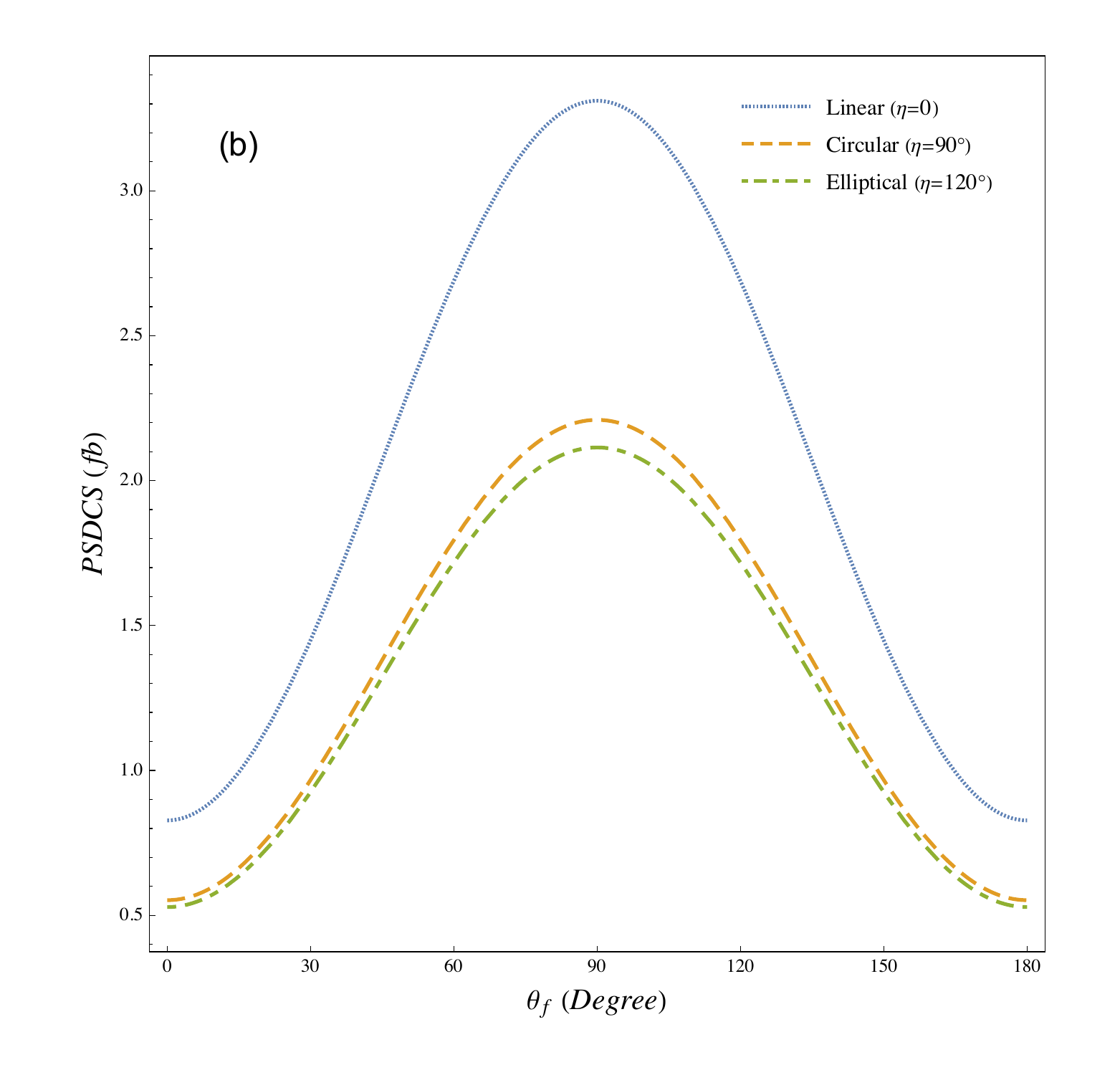}\hspace*{0.1cm}
		\caption{ The PSDCS as a function of outgoing polar angle $\theta_{f}$. The laser field strength and frequency are fixed at $\varepsilon=10^{5} \,\, \mathrm{V.cm^{-1}}$ and $\omega=1.17 \,\, \mathrm{eV}$, respectively, with $-3 \leq n \leq 3$. The doubly charged Higgs boson mass and the centre-of-mass energy are fixed at $M_{H^{\pm\pm}}=300\,\mathrm{GeV}$ and $\sqrt{s}=1000\,\mathrm{GeV}$, respectively.  The scattering geometry is defined by $\phi_{f}=\phi_{i}+90^{\circ}$, with (a): $\theta_{i}=\phi_{i}= 6^{\circ}$ and in (b): $\theta_{i}=\phi_{i}= 30^{\circ}$.}
		\label{fig5}
	\end{figure}

\begin{figure}[H]
		\centering
		\includegraphics[scale=0.45]{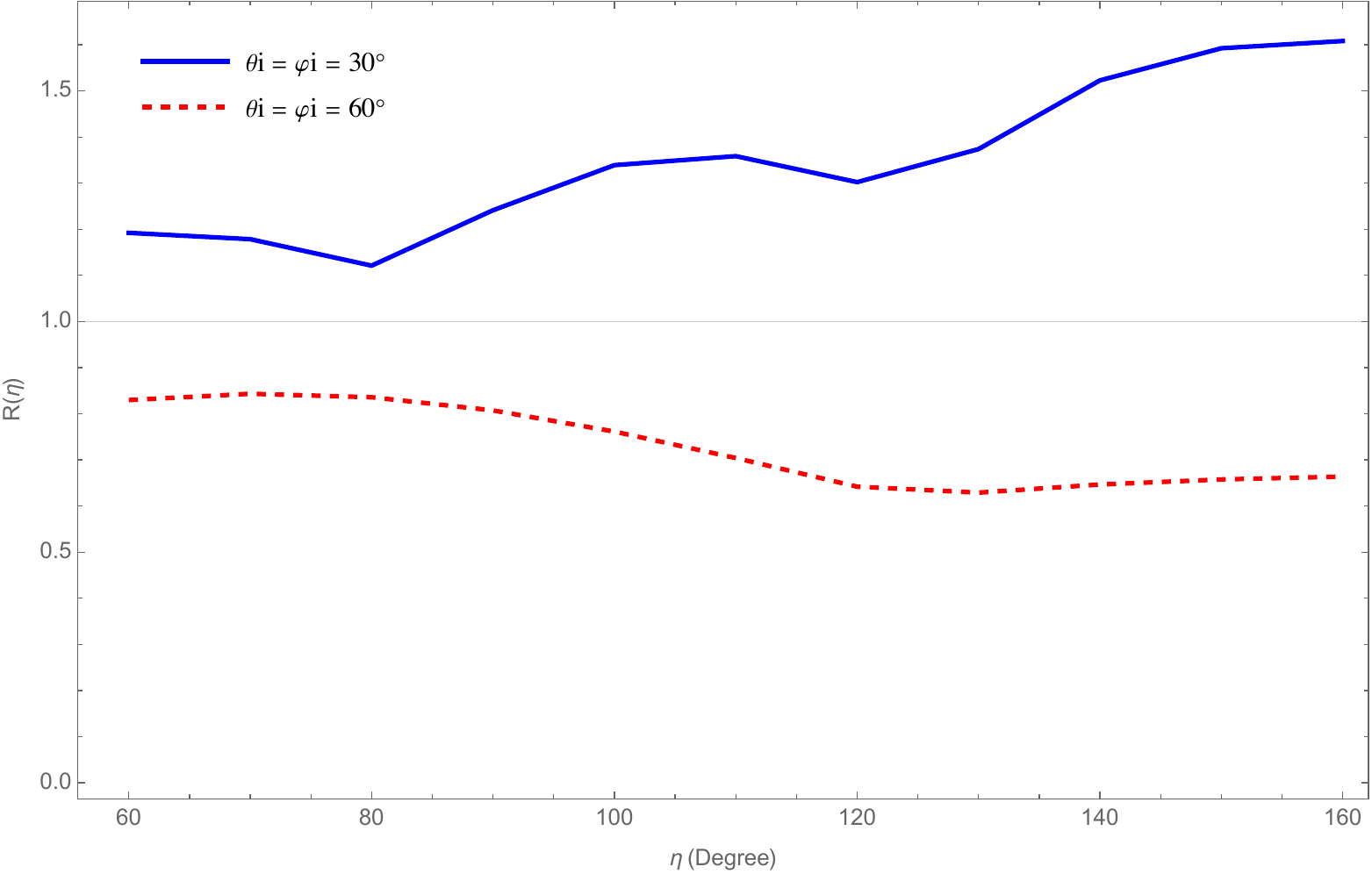}
		\caption{The ratio $R(\eta)$ of the maximum harmonic-resolved DCS at a given ellipticity to that obtained for linear polarization, as a function of the ellipticity parameter $\eta$, for two incidence geometries: $\theta_{i}= \phi_{i}= 30^{\circ}$ and $60^{\circ}$. The reference laser field strength and frequency are fixed at $\varepsilon=10^{5} \,\,\mathrm{V.cm^{-1}}$ and $\omega=1.17 \,\,\mathrm{eV}$, respectively. The doubly charged Higgs boson mass and the centre-of-mass energy are fixed at $M_{H^{\pm\pm}}=300\,\mathrm{GeV}$ and $\sqrt{s}=1000\,\mathrm{GeV}$, respectively. $\theta_{f}=90^{\circ}$ and $\phi_{f}= \phi_{i} + 90^{\circ}$.}
		\label{fig6}
	\end{figure}

\begin{figure}[H]
		\centering
		\includegraphics[scale=0.35]{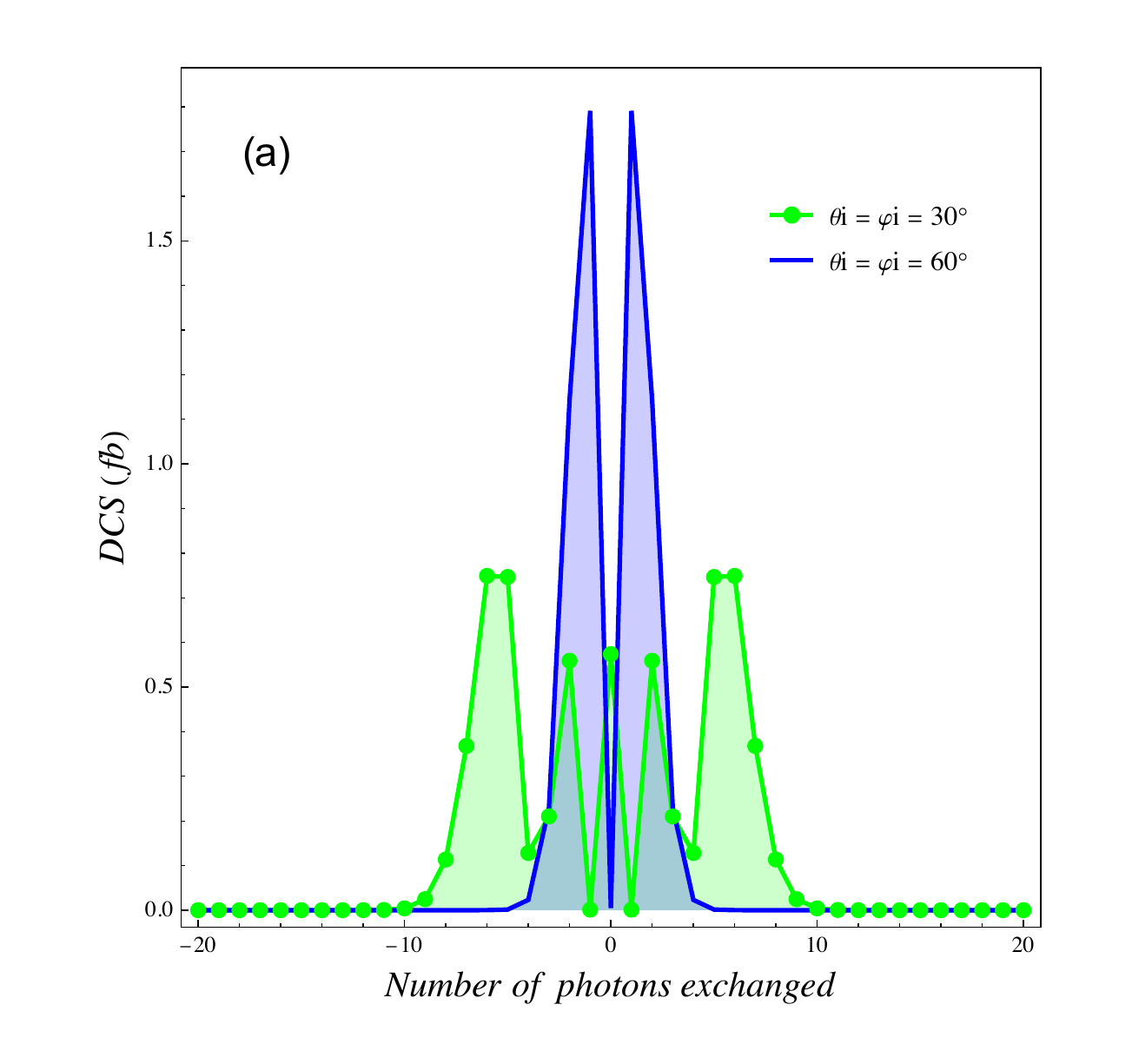}\hspace*{0.1cm}
		\includegraphics[scale=0.35]{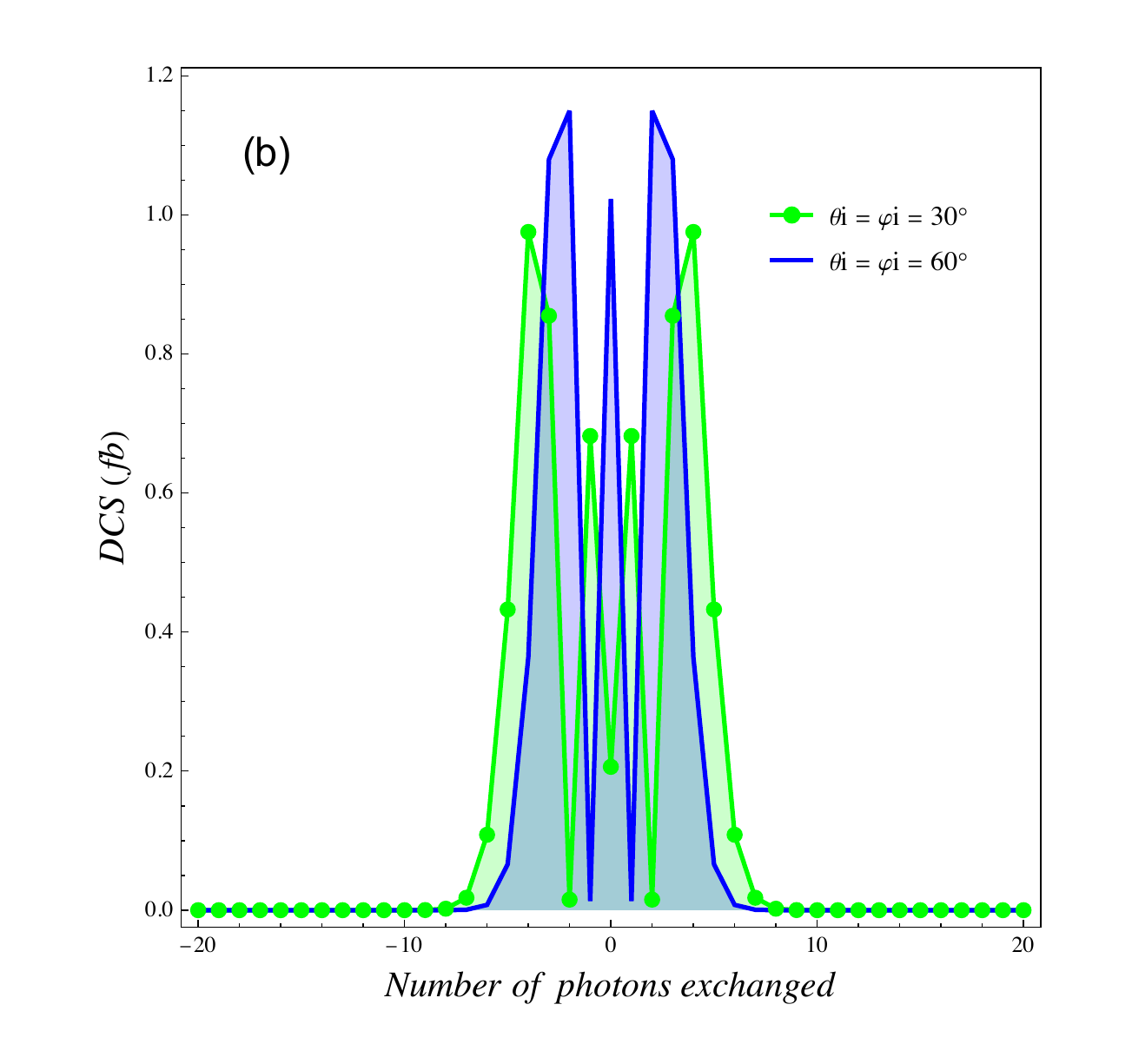}\hspace*{0.1cm}
		\caption{ The DCS as a function of the number of photons exchanged for two incidence geometries: $\theta_{i}=\phi_{i}= 30^{\circ}$ and $60^{\circ}$, considering (a): Linear and (b): elliptical ($\eta=120^{\circ}$) polarization. The reference laser field strength and frequency are fixed at $\epsilon_{0}=10^{5}\,\mathrm{V\,cm^{-1}}$ and $\omega=1.17\,\mathrm{eV}$, with the same fixed average potential normalization as in Fig.~\ref{fig6}. The doubly charged Higgs boson mass and the centre-of-mass energy are fixed at $M_{H^{\pm\pm}}=300\,\mathrm{GeV}$, and $\sqrt{s}=1000\,\mathrm{GeV}$, respectively.}
		\label{fig7}
	\end{figure}

\begin{figure}[H]
		\centering
		\includegraphics[width=0.31\textwidth]{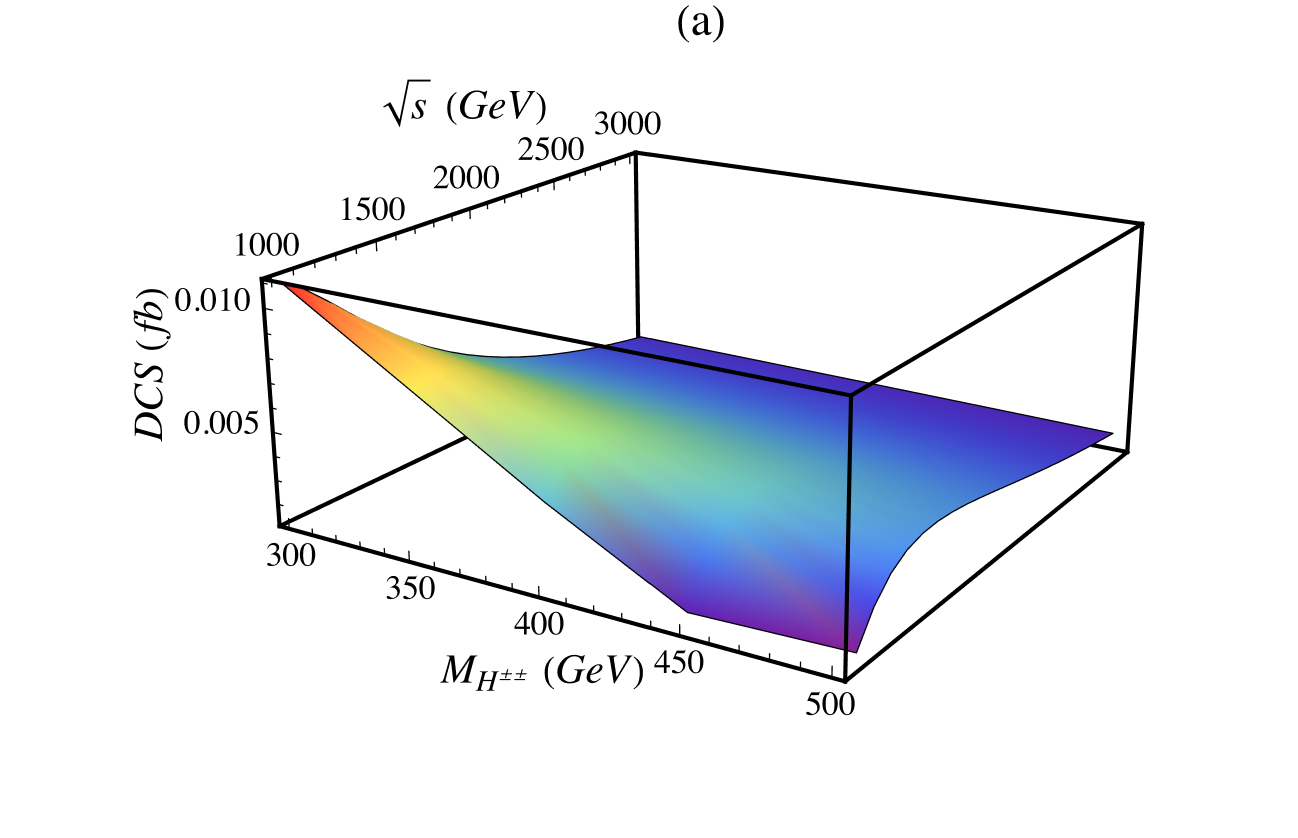}\hfill
		\includegraphics[width=0.31\textwidth]{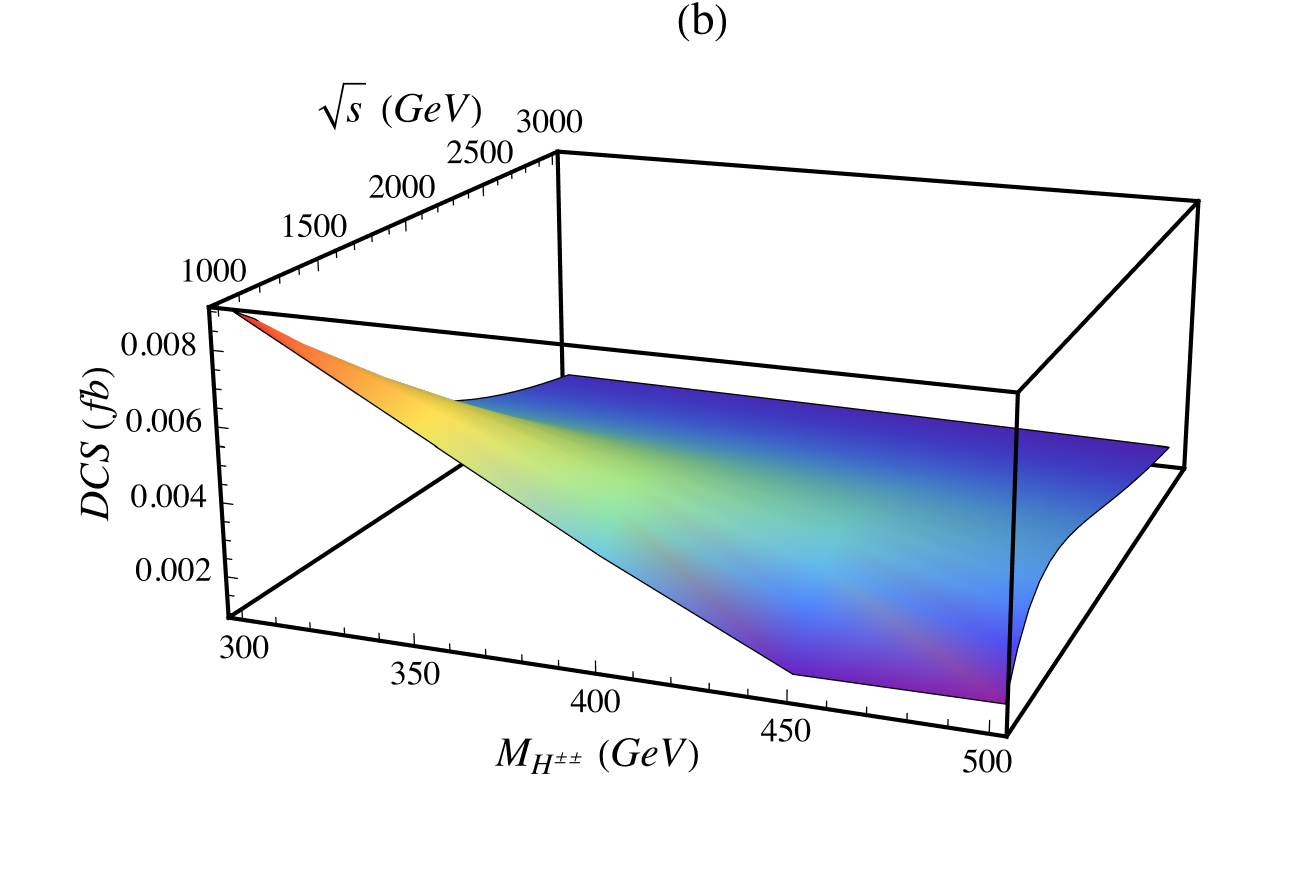}\hfill
		\includegraphics[width=0.31\textwidth]{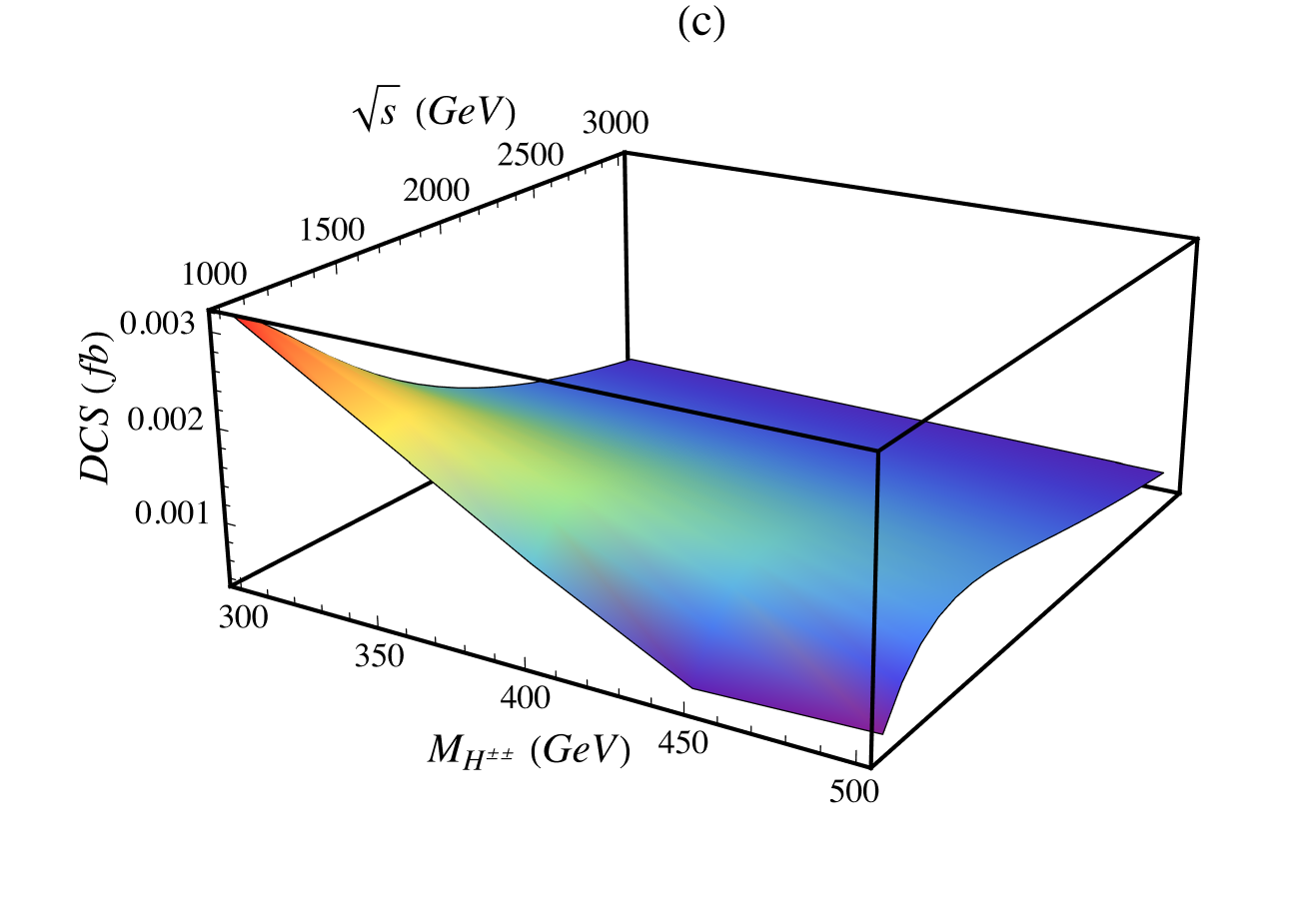}
		
		\vspace{0.3cm}
		
		\includegraphics[width=0.31\textwidth]{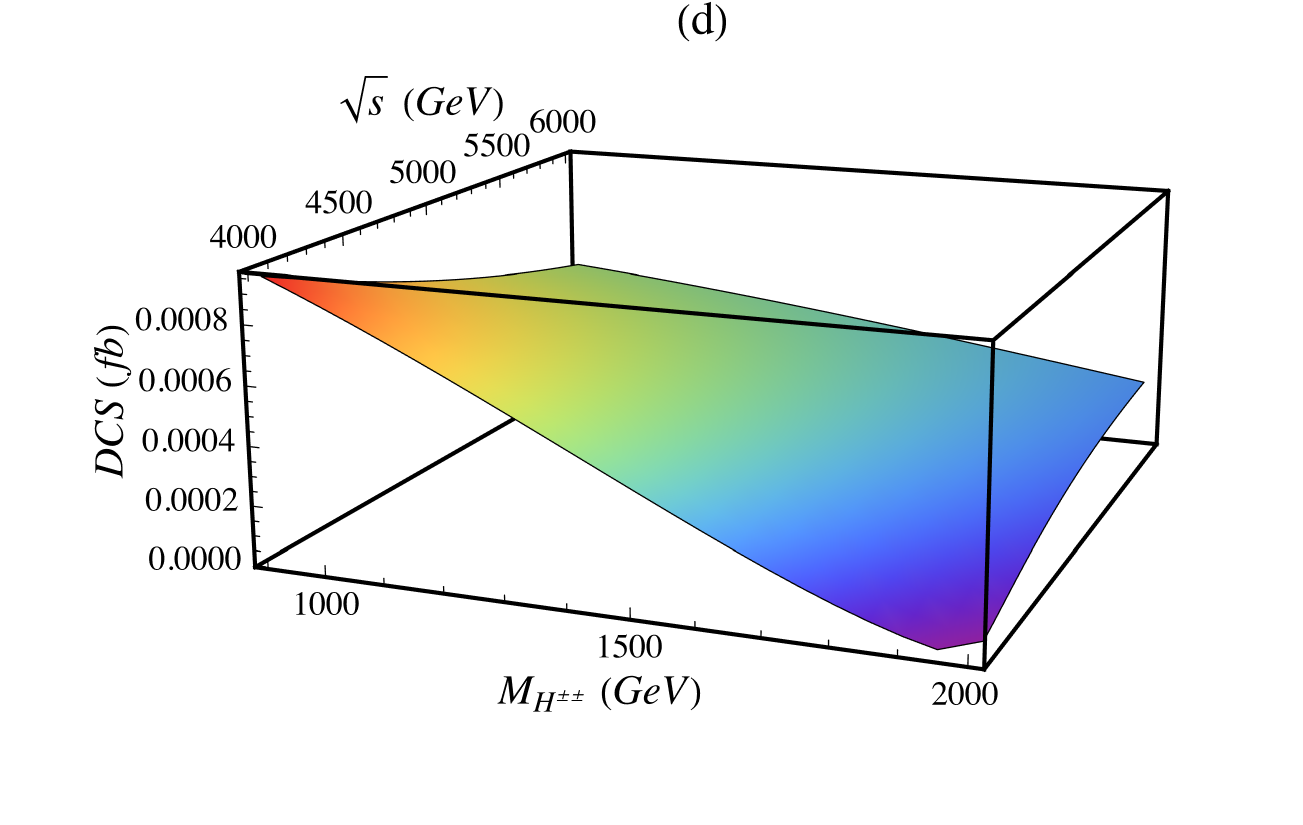}\hfill
		\includegraphics[width=0.31\textwidth]{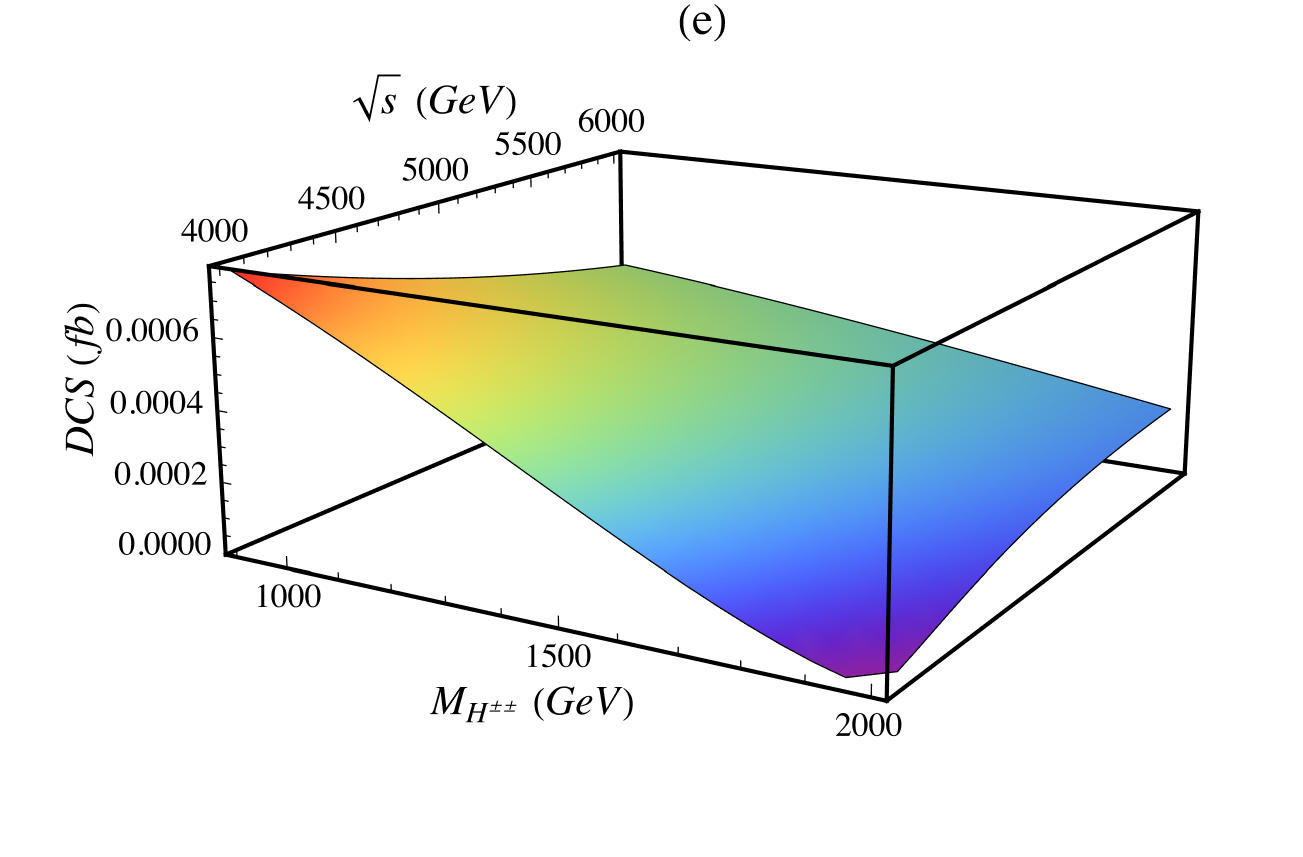}\hfill
		\includegraphics[width=0.31\textwidth]{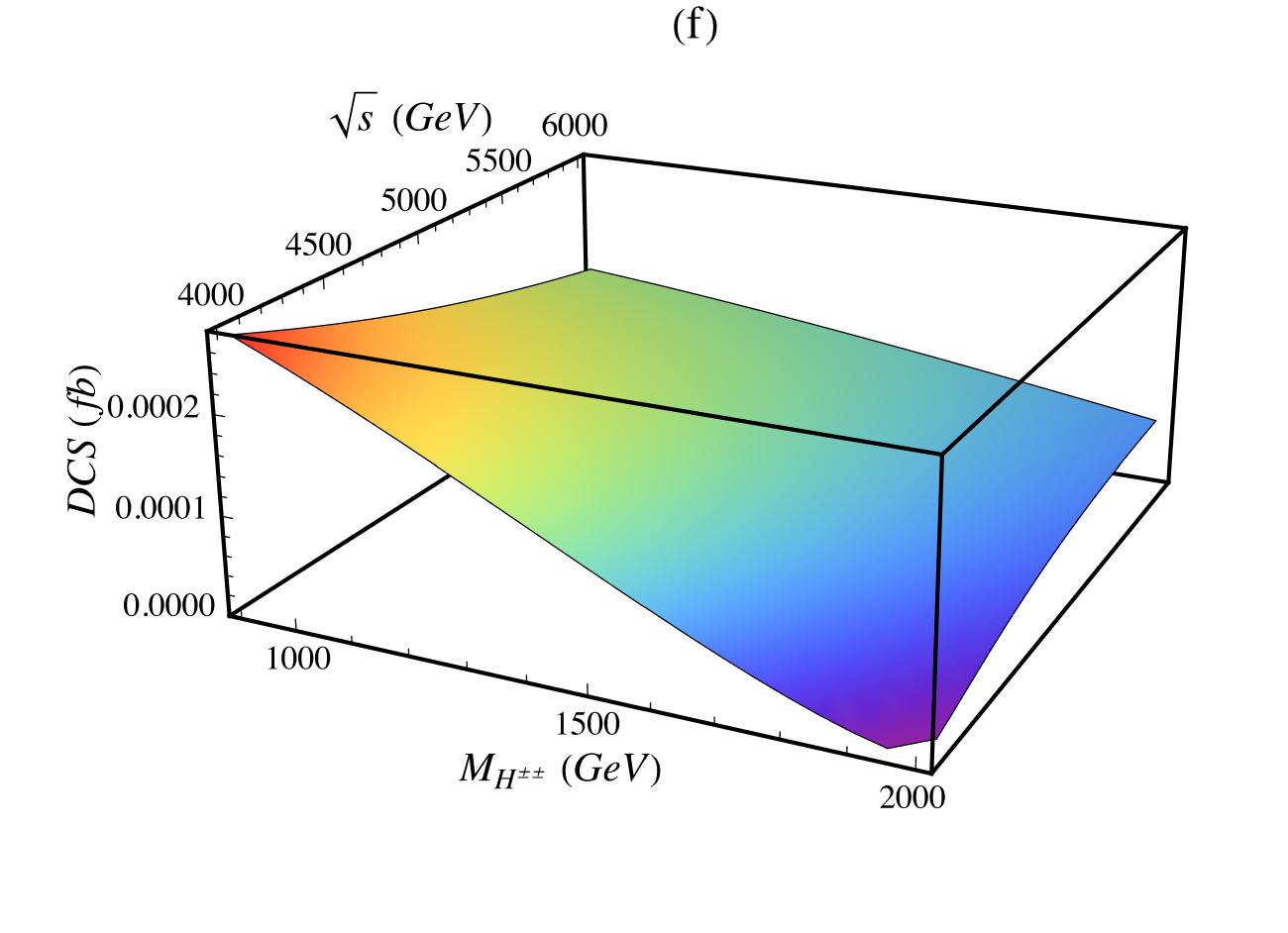}
		\caption{The DCS as a function of $\sqrt{s}$ and  $M_{H^{\pm\pm}}$ for a reference laser field strength $\varepsilon=10^{7} \,\,V.cm^{-1}$, a frequency $\omega=2\,\,eV$  and a fixed number of photons exchanged $n=160$. The scattering geometry is defined by $\theta_{i}=\phi_{i}=30^{\circ}$, $\theta_{f}=90^{\circ}$ and $\phi_{f}=\phi_{i}+90^{\circ}$. Panels (a,d), (b,e) and (c,f) correspond to linear $\eta=0$, circular $\eta=90°$, and elliptical $\eta=120^{\circ}$ polarization, respectively. Panels (a), (b) and (c) cover the range  $300\leq M_{H^{\pm\pm}} \leq  
			500\,\, GeV$ and $1000 \leq \sqrt{s} \leq3000\,\, GeV$, while panels (d), (e) and (f) cover  $900 \leq M_{H^{\pm\pm}} \leq 2000\,\, GeV$, $4000 \leq \sqrt{s} \leq 6000\,\, GeV$.  }
		\label{fig8}
	\end{figure}

	In the presence of an external laser field, the squared transition amplitude $\big|\overline{M}_{fi}\big|^2$ involves several lengthy Dirac traces containing ordinary and generalized Bessel functions. Consequently, the evaluation of the TCS, which requires both angular integration and summation over the exchanged photon number, becomes computationally demanding. To investigate more directly the relationship between laser polarization and scattering geometry, we therefore focus on the DCS and on its summation over $n$ photons exchanged. When the summation is restricted to a finite interval, we refer to the resulting quantity as the partially summed differential cross section PSDCS:
	\begin{equation}
		PSDCS=\sum_{n=-N}^{N} \dfrac{d\sigma_{n}}{d\Omega}
	\end{equation}
	In Fig.~\ref{fig3}, we plot the PSDCS, evaluated over the number of photons exchanged interval $-3 \leq n \leq 3$, as a function of the scattering angle $\Delta\theta=\phi_{f} -\phi_{i}$ for $\theta_{i}= 90^{\circ}$, $\phi_{i}= 0^{\circ}$ and $\theta_{f}=90^{\circ}$. The small difference from the laser-free result should also be interpreted in the context of the finite photon exchange interval considered here.\\
	The curves corresponding to the different polarization states overlap within numerical accuracy, indicating that the polarization dependence is strongly suppressed for this particular geometry.\\
	This behavior can be explained by the structure of the Bessel arguments. For $\theta_{i}= 90^{\circ}$ and $\phi_{i}= 0^{\circ}$, the incoming particles momentum lie along the $(Ox)$ direction, while the second polarization four-vector $a_{2}^{\mu}$ is directed along $(Oy)$. Hence, $a_{2}.p_{1}=a_{2}.p_{2}=0$, which implies $\alpha_2=0$. The Bessel argument therefore reduces to $z=\big|\alpha_1\big|$ and becomes independent of the ellipticity parameter. In addition, for the relatively weak field strength considered here, the remaining ellipticity-dependent contribution through $\alpha_3$ is numerically very small. As a result, the PSDCS becomes nearly insensitive to the polarization state in this specific geometry.
	However, this situation changes when the azimuthal angle for the incoming particles is varied. In Fig.~\ref{fig4}-(a), we set $\phi_{i}=30^{\circ}$, while keeping the same number of photons exchanged interval $-3 \leq n \leq 3$. The curves corresponding to the different polarization states are no longer identical, revealing the sensitivity of the PSDCS to the chosen scattering geometry.\\
	This observation arises from the dependence of the Bessel arguments on the projections of the incoming particle momentum onto the two polarization directions. For $\theta_{i}= 90^{\circ}$, the quantities entering $\alpha_{1}$ and $\alpha_{2}$ are proportional to the $Ox$ and $Oy$ components of the incoming momentum, respectively. Consequently, their azimuthal dependence is governed by $\cos \phi_{i}$ and $\sin \phi_{i}$. For linear polarization, $\tan(\dfrac{\eta}{2})=0$, so that $\alpha_{2}=0$, and the variation with $\phi_{i}$ arises mainly through $\alpha_{1}$. In contrast, for circular polarization $\tan(\dfrac{\eta}{2})=1$, and the two transverse components enter symmetrically. As a consequence, the combination $z=\sqrt{ \alpha_{1}^2+\alpha_{2}^2}$ becomes insensitive to a rotation of the incoming momentum in the transverse plane, which is consistent with the weak variation of the circular polarization result with $ \phi_{i}$.\\
	When the azimuthal angle is further increased to $\phi_{i}=60^{\circ}$, as shown in Fig.~\ref{fig4}-(b), the separation between the polarization configurations becomes more pronounced, particularly for elliptical polarization. In this case, $\tan(\dfrac{\eta}{2}) \neq 1$ and the two transverse momentum components enter the Bessel argument with different weights. The modification of $\phi_{i}$ therefore changes the multiphoton coefficients and leads to a stronger geometry-dependent modification of the PSDCS. \\
	Fig.~\ref{fig4} thus shows that the sensitivity of the PSDCS to the laser polarization is not universal, but is controlled by the relative orientation of the incoming particle momenta and the polarization of the external field. This provides a first clear indication of the close connection between polarization and scattering geometry.\\
	It is important to emphasize that the same finite photon number interval ($-3 \leq n \leq 3$) is maintained for all three scattering geometries. Therefore, the increasing separation between the polarization configurations reflects the geometry-dependent redistribution of multiphoton contributions within the selected harmonic range, rather than a corresponding suppression of the fully inclusive cross section.
	To extend this analysis, we now vary the incident angles $\theta_{i}$ and $\phi_{i}$. The corresponding results are displayed in Fig.~\ref{fig5}, where the PSDCS is plotted as a function of the outgoing polar angle $\theta_{f}$. As shown in Fig.~\ref{fig5}-(a), for small angles $\theta_{i}=\phi_{i}= 6^{\circ}$, corresponding to a nearly aligned configuration between the incident beam axis and the laser propagation direction, the value of the PSDCS is considerably reduced within the selected photon exchange interval. In this configuration, the curves associated with the different polarization states remain very close to each other, indicating that the relative influence of polarization is weak. This suggests that, for nearly aligned geometry, the sensitivity of the PSDCS to the laser polarization becomes less pronounced. By contrast, for $\theta_{i}=\phi_{i}= 30^{\circ}$, shown in Fig.~\ref{fig5}-(b), the PSDCS curves for various polarizations become clearly distinguishable. In this regime, the effect of the laser polarization becomes more visible, with clear distinctions between linear, circular, and elliptical polarization states. \\
	These results further demonstrate that the influence of laser polarization on the multiphoton contributions depends strongly on the scattering geometry, consistent with the behavior observed in Fig.~\ref{fig4}.
	To further quantify the interplay between laser polarization and scattering geometry, we investigate the dependence of the dominant harmonic contribution on the ellipticity parameter $\eta$. While Fig.~\ref{fig3} to Fig.~\ref{fig5} illustrate the polarization dependence of the partially summed differential cross section within a fixed photon-exchange interval, we now examine the maximum contribution of the harmonic-resolved DCS.\\
	For this purpose, we introduce the ratio $R(\eta)$ defined as the ratio between the maximum harmonic DCS at a given ellipticity and the corresponding maximum obtained for linear polarization under the same scattering geometry. The representative harmonic order is determined separately for each polarization state by searching over the interval $-40 \leq n \leq 40$.\\
	To disentangle the effect of ellipticity from the variation of the average squared potential, the laser four-potential is rescaled by the factor:
	\begin{equation}
		\sqrt{\dfrac{2}{1+\tan^{2}(\eta/2)}}
	\end{equation}
	so that $\bar{A^{2}}$ remains constant for all values of $\eta$. The linear polarization ($\eta=0^{\circ}$) is therefore used as a common reference with $R(0^{\circ})=1$.\\
	As illustrated in Fig.~\ref{fig6}, the two scattering geometries respond differently to laser ellipticity. For $\theta_{i}=\phi_{i}= 30^{\circ}$, the ratio exceeds unity throughout the investigated ellipticity range, indicating an enhancement of the dominant harmonic contribution relative to the linear polarization case. In particular, at $\eta=120^{\circ}$, the ratio reaches approximately $1.30$, corresponding to an increase of about $30\%$.\\
	By contrast, for $\theta_{i}=\phi_{i}= 60^{\circ}$, the ratio remains below unity and generally decreases as the ellipticity increases. Thus, for this geometry, the dominant harmonic contribution is reduced relative to the linear polarization case.\\
	These contrasting behaviors reveal a pronounced geometry dependence of the response to ellipticity. In particular, elliptical polarization can either enhance or suppress the dominant harmonic contribution depending on the orientation of the incident particle momentum. The persistence of this trend under a fixed cycle-averaged squared potential indicates that the observed differences cannot be attributed solely to variations in the average field normalization.\\
	Beyond extending the formalism from circular to general elliptical polarization, the present analysis identifies a geometry-dependent reversal in the relative strength of the dominant harmonic contributions.\\
	It should be emphasized, however, that the present ratio characterizes the maximum harmonic contribution within the selected photon number interval and does not necessarily imply a corresponding enhancement or suppression of the fully photon-summed cross section.\\
	To gain deeper insight into the geometry-dependent behavior of the peak ratio observed in Fig.~\ref{fig6}, we present in Fig.~\ref{fig7} the harmonic-resolved DCS as a function of the number of exchanged photons for two incidence geometries, $\theta_{i}= \phi_{i}= 30^{\circ}$ and $60^{\circ}$, considering linear and elliptical laser polarizations. As in Fig.~\ref{fig6}, the cycle-averaged squared laser potential is kept constant.\\
	For both polarization states, increasing the incidence angles from $30^{\circ}$ to $60^{\circ}$ leads to a narrower harmonic distribution, with the dominant contributions concentrated around lower photon orders. However, the relative heights of the harmonic peaks exhibit a pronounced polarization dependence.\\
	In particular, for $\theta_{i}= \phi_{i}= 30^{\circ}$, the maximum harmonic DCS obtained with elliptical polarization is approximately $30.22\%$ higher than that obtained with linear polarization, corresponding to $R(\eta=120^{\circ})=1.30222$. In contrast, for $\theta_{i}= \phi_{i}= 60^{\circ}$, the elliptical polarization is reduced by approximately $35.81\%$ relative to the linear case, yielding $R(\eta=120^{\circ})=0.641882$.\\
	These results provide a direct interpretation of the opposite trends observed in Fig.~\ref{fig6}. They demonstrate that the scattering geometry can reverse the relative hierarchy of the dominant harmonic contributions under different laser polarizations, even when the average squared laser potential is maintained constant.\\
	For the laser-assisted process ${e}^{+}{e}^{-}\rightarrow H^{++}H^{--}$, this finding indicates that the contributions of individual photon-exchange channels depend on the combined effects of laser ellipticity and the orientation of the incident particle momentum. The polarization dependence therefore cannot be characterized independently of the scattering geometry.\\
	To complete the analysis, we now investigate the dependence of the harmonic-resolved differential cross section on both the centre-of-mass energy $\sqrt{s}$ and the doubly charged Higgs boson mass $M_{H^{\pm\pm}}$,  considering linear, circular, and elliptical laser polarizations. We also explore a higher-mass region motivated by recent collider searches. The experimental lower bounds on $M_{H^{\pm\pm}}$ depend strongly on the underlying model, the assumed decay modes, and the corresponding branching ratios. In same-sign leptonic searches, the ATLAS Collaboration \cite{45} obtained an observed lower mass limit of $1080\,\, GeV$ within the left-right symmetric type-II seesaw interpretation, while a lower limit of $900\,\, GeV$ was obtained in the Zee-Babu model. These results illustrate that current direct searches probe doubly charged scalars at the TeV scale, although the precise limits are model-dependent. Motivated by this mass scale, we therefore extend the analysis to the range $900 \leq M_{H^{\pm\pm}} \leq 2000 \,\, GeV $ with $4000 \leq \sqrt{s} \leq 6000\,\, GeV$.\\
	The calculations are performed at a fixed photon number exchanged $n=160$, with a reference field strength $\epsilon_{0}=10^{7}\,\mathrm{V\,cm^{-1}}$ and $\omega=2\,\mathrm{eV}$, while maintaining the same cycle-averaged squared laser potential for all polarization states.\\
	The results in Fig.~\ref{fig8} show that the DCS depends strongly on the production kinematics, particularly as the doubly charged Higgs boson mass approaches the kinematic threshold. Moreover, a clear polarization hierarchy is observed, with the linear polarization producing the largest contribution, followed by circular and elliptical polarizations. This ordering was verified numerically at $48$ kinematically allowed points distributed across the two investigated regions, with no reversal observed.\\
	While the analysis presented in Fig.~\ref{fig6} and Fig.~\ref{fig7}, the comparison here is performed for the same number of photons exchanged. The results therefore allow us to investigate the dependence of the DCS on the collision energy and Higgs boson mass, providing a complementary characterization of the laser-assisted production process.
	However, the characteristic multiphoton domain itself changes significantly with the scattering geometry. To investigate this dependence more directly, we vary the incidence angles and determine, for each polarization state, a representative number of photons exchanged $n_{rep}$ identified as the photon order at which the DCS reaches its largest value within the investigated positive-harmonic region around the corresponding Bessel argument $z$. Unlike the analysis in Fig.~\ref{fig8}, where the number of photons exchanged is fixed at $n=160$, this procedure allows the dominant harmonic contribution to be examined separately for each polarization and incidence geometry.\\
	Two mass-energy configurations are considered: $(M_{H^{\pm\pm}},\sqrt{s})=(400\,\, GeV, 1600\,\, GeV) $ and $(1500\,\, GeV, 6000\,\, GeV)$. Both satisfy the same nominal ratio $\sqrt{s}/(2M_{H^{\pm\pm}})=2$, providing a common kinematic basis for comparing geometry and polarization effects at two different mass scales. The laser parameters are kept identical to those used in Fig.~\ref{fig8}, with $\varepsilon=10^{7} \,\,V.cm^{-1}$ and $\omega=2\,\,eV$, while maintaining a constant cycle-averaged squared laser potential across the polarization states.\\
	Tables~\ref{tab:Table1} and \ref{tab:Table2} report the Bessel argument $z$, the representative number of photons exchanged $n_{rep}$, and the corresponding DCS for $\theta_{i}=\phi_{i}=6^{\circ}, 30^{\circ}, 60^{\circ}$, considering linear, circular and elliptical polarization. A first important feature is that $n_{rep}$ closely follows the value of $z$, indicating that the Bessel argument provides a useful estimate of the harmonic region where the largest positive-harmonic contributions are found.\\
	A more significant result emerges from the dependence of the representative harmonic contributions on the incidence geometry. As the incidence angles increase from $6^{\circ}$ to $60^{\circ}$, the Bessel argument $z$ decreases for all three polarization states, accompanied by a shift of the representative photon-exchange number toward lower values. This behavior indicates that the scattering geometry strongly influences the harmonic region contributing most significantly to the laser-assisted production of doubly charged Higgs boson pairs.\\
	Interestingly, the corresponding representative DCS increases with the incidence angles for all polarization configurations. However, the relative contributions of the different polarizations exhibit a pronounced geometry dependence. For $\theta_{i}=\phi_{i}=6^{\circ}$ and $30^{\circ}$, the elliptical polarization provides the largest representative DCS, followed by the circular and linear cases. In contrast, at $\theta_{i}=\phi_{i}=60^{\circ}$, this hierarchy is reversed, with linear polarization yielding the highest value.\\
	The observed polarization hierarchy can be understood in terms of the interaction between the laser polarization and the geometry-dependent Volkov phases. The projections of the incident particle momenta onto the laser polarization directions determine the Bessel arguments and modify the harmonic coefficients entering the scattering amplitude. Consequently, changing the incidence geometry affects both the number of photons exchanged at which the dominant contributions occur and their relative magnitudes. Since the cycle-averaged squared laser potential is kept constant, these differences cannot be attributed simply to a change in the average laser intensity.\\
	For the process ${e}^{+}{e}^{-}\rightarrow H^{++}H^{--}$, these results indicate that the relative strengths of the dominant laser-assisted photon-exchange channels can be enhanced or suppressed by changing the incidence geometry and polarization. The reversal observed between $30^{\circ}$ and $60^{\circ}$ is consistent with the geometry-dependent behavior identified in Fig.~\ref{fig6} and  Fig.~\ref{fig7}, despite the different laser parameters considered here.\\
	Furthermore, both mass-energy configurations exhibit the same representative harmonic numbers and polarization hierarchy, while the absolute DCS values decrease at the larger mass-energy scale. This suggests that, for the two benchmarks considered at fixed $\sqrt{s}/(2M_{H^{\pm\pm}})$, the geometry-dependent redistribution of the dominant harmonic contributions remains largely unchanged.\\ 
	\begin{table}[H]
		\centering
		\small
		\setlength{\tabcolsep}{3pt}
		\begin{minipage}{0.48\textwidth}
			\centering
			\refstepcounter{table}
			\label{tab:Table1}
			\textbf{Table \thetable}\par\vspace{2mm}
			
			\resizebox{\linewidth}{!}{%
				\begin{tabular}{|c|c|c|c|c|}
					\hline
					$\theta _i = \varphi _i (^{\circ})$ & Polarization & $z$ & $n_{rep}$ & $DCS(n_{rep})$  (fb) \\
					\hline
					6 & Linear & 1327.55 & 1319 & 0.0119152 \\
					\hline
					6 & Circular & 943.891 & 936 & 0.0149867 \\
					\hline
					6 & Elliptical ($120^{\circ}$) & 674.685 & 668 & 0.0186914 \\
					\hline
					30 & Linear & 241.675 & 237 & 0.0370837 \\
					\hline
					30 & Circular & 197.327 & 193 & 0.0424208 \\
					\hline
					30 & Elliptical ($120^{\circ}$) & 170.89 & 166 & 0.0466635 \\
					\hline
					60 & Linear & 80.5584 & 77 & 0.0780306 \\
					\hline
					60 & Circular & 113.927 & 110 & 0.0617966 \\
					\hline
					60 & Elliptical ($120^{\circ}$) & 127.374 & 123 & 0.057004 \\
					\hline
				\end{tabular}%
			}
		\end{minipage}
		\hfill
		\begin{minipage}{0.48\textwidth}
			\centering
			\refstepcounter{table}
			\label{tab:Table2}
			\textbf{Table \thetable}\par\vspace{2mm}
			
			\resizebox{\linewidth}{!}{%
				\begin{tabular}{|c|c|c|c|c|}
					\hline
					$\theta _i = \varphi _i (^{\circ})$ & Polarization & $z$ & $n_{rep}$ & $DCS(n_{rep})$  (fb) \\
					\hline
					6 & Linear & 1327.55 & 1319 & 0.000846567 \\
					\hline
					6 &Circular & 943.891 & 936 & 0.0010648 \\
					\hline
					6 & Elliptical ($120^{\circ}$)  &  674.685 & 668 & 0.00132801 \\
					\hline
					30 & Linear & 241.675 & 237 & 0.00263478 \\
					\hline
					30 & Circular & 197.327 & 193 & 0.00301397 \\
					\hline
					30 & Elliptical ($120^{\circ}$)  & 170.89 & 166 & 0.00331542 \\
					\hline
					60 & Linear & 80.5584 & 77 & 0.00554403 \\
					\hline
					60 & Circular & 113.927 & 110 & 0.00439062 \\
					\hline
					60 & Elliptical ($120^{\circ}$) & 127.374 & 123 & 0.0040501 \\
					\hline
				\end{tabular}%
			}
		\end{minipage}
		\caption*{\textbf{Tables \ref{tab:Table1} and \ref{tab:Table2}:}
			Bessel argument $z$, representative number of photons exchanged $n_{\mathrm{rep}}$, and corresponding harmonic-resolved DCS for three incidence geometries and laser polarization states. The reference laser field strength and frequency are $\epsilon_{0}=10^{7}\,\mathrm{V\,cm^{-1}}$ and $\omega=2\,\mathrm{eV}$, respectively, with $\theta_{f}=90^{\circ}$ and $\phi_{f}=\phi_{i}+90^{\circ}$. Table \ref{tab:Table1} corresponds to $M_{H^{\pm\pm}}=400\,\, GeV$ and $\sqrt{s}= 1600\,\, GeV $, while Table \ref{tab:Table2} corresponds to $M_{H^{\pm\pm}}=1500\,\, GeV$ and $\sqrt{s}= 6000\,\, GeV $.}
	\end{table}

	\section{Conclusion}\label{concl}
	In this work, we have established the analytical expression for the DCS of laser-assisted doubly charged Higgs boson pair production through electron-positron annihilation in the HTM using a general elliptically polarized monochromatic plane wave. The formalism provides a unified description of linear, circular, and elliptical polarization states and has been validated against previous results obtained in the circular polarization limit.\\
	Our analysis demonstrates that the effect of laser polarization cannot be characterized independently of the scattering geometry. For nearly collinear incidence, the polarization dependence becomes weak, whereas other geometrical configurations reveal significant differences between polarization states. In particular, the peak ratio normalized to linear polarization exhibits opposite ellipticity-dependent trends for different incidence geometries. A notable result is the geometry-dependent reversal of the relative strengths of the dominant harmonic contributions, showing that elliptical polarization can enhance or suppress these contributions relative to linear polarization depending on the incidence geometry. \\
	Furthermore, the analysis of the centre-of-mass energy and the doubly charged Higgs boson mass reveals a strong kinematic dependence of the DCS. The representative harmonic analysis also shows that the geometry-dependent polarization hierarchy is preserved at the two mass-energy scales investigated, although the magnitude of the DCS decreases at higher energies and masses.\\
	Overall, these findings highlight the scattering geometry as a key factor controlling the polarization sensitivity and redistribution of multiphoton contributions in laser-assisted $e^{+}e^{-}\rightarrow H^{++}H^{--}$ production. 
	       
\end{document}